\documentclass{emulateapj}
\usepackage{amsmath}
\usepackage{graphicx}
\usepackage{amsfonts}
\usepackage{natbib}
\usepackage{overpic}

\newcommand{\be}{\begin{equation}}
\newcommand{\ee}{\end{equation}}
\newcommand{\ba}{\begin{align}}
\newcommand{\ea}{\end{align}}

\newcommand{\si}{\sigma}

\newcommand{\ka}{\kappa}
\newcommand{\lam}{\lambda}

\newcommand{\scalefit}{{\texttt{ScaleFit}}}
\newcommand{\boxfit}{{\texttt{BoxFit}}}
\newcommand{\ramcode}{{\texttt{RAM}}}
\newcommand{\blastcode}{{\texttt{BLAST}}}
\newcommand{\swift}{{\it Swift}}
\newcommand{\swiftXRT}{{\it Swift-XRT}}
\newcommand{\vela}{{\it Vela}}
\newcommand{\bepposax}{{\it BeppoSAX}}
\newcommand{\chandra}{{\it Chandra}}

\newcommand{\emcee}{{\texttt{emcee}}}
\newcommand{\multinest}{\texttt{MultiNest}}
\newcommand{\afterglowlibrary}{{\texttt{http://cosmo.nyu.edu/afterglowlibrary/}}}

\newcommand{\Eiso}{E_{iso}}
\newcommand{\thO}{\theta_0}
\newcommand{\thobs}{\theta_{obs}}
\newcommand{\epse}{\epsilon_e}
\newcommand{\epsB}{\epsilon_B}
\newcommand{\xN}{\xi_N}
\newcommand{\tobs}{t_{obs}}
\newcommand{\Fp}{F_{peak}}
\newcommand{\cf}{\mathfrak{f}}
\newcommand{\cfp}{\cf_{peak}}
\newcommand{\cfm}{\cf_{m}}
\newcommand{\cfc}{\cf_{c}}

\begin{document}

\title{Gamma Ray Bursts Are Observed Off-Axis}
\author{Geoffrey Ryan\altaffilmark{1,a}, Hendrik van Eerten\altaffilmark{1,2}, Andrew MacFadyen\altaffilmark{1}, and Bin-Bin Zhang\altaffilmark{3}}
\altaffiltext{a}{gsr257@nyu.edu}
\altaffiltext{1}{Center for Cosmology and Particle Physics, Physics Department, New York University, NY, NY 10003, USA}
\altaffiltext{2}{Alexander von Humboldt Fellow
Max-Planck Institute for Extraterrestrial Physics (MPE)
Postfach 1312, 85741 Garching, Germany}
\altaffiltext{3}{Center for Space Plasma and Aeronomic Research (CSPAR), University of Alabama in Huntsville, 
Huntsville, AL 35899, USA}

\begin{abstract}

We use the \scalefit{} package to perform Markov chain Monte Carlo light curve fitting on a large sample of \swiftXRT{} gamma-ray burst afterglows.  The \scalefit{} model uses scaling relations in the hydrodynamic and radiation equations to compute synthetic light curves directly from a set of high resolution two-dimensional relativistic blast wave simulations.  The data sample consists of all \swiftXRT{} afterglows from 2005 to 2012 with sufficient coverage and a known redshift, 188 bursts in total.  We find the jet half opening angle varies widely but is commonly less than 0.1 radians.  The distribution of the electron spectral index is also broad, with a median at 2.23.  This approach allows, for the first time, for the off-axis observer angle to be inferred directly from the light curve.  We find the observer angle to have a median value of 0.57 of the jet opening angle over our sample, which has profound consequences for the predicted rate of observed jet breaks and effects the beaming corrected total energies of gamma-ray bursts.

\end{abstract}


\section{Introduction}

Gamma-ray bursts (GRBs) are intense flashes of gamma-rays believed to be caused by collapsing massive stars and merging compact binaries \citep{Woosley93, MacFadyen99, Eichler89}.  First observed by the \vela{} satellites in 1969, GRBs occur isotropically in the sky and persist for as long as several minutes or as short as fractions of a second \citep{Klebesadel73}.  In 1997 the \bepposax{} satellite detected a faint, decaying x-ray signal from GRB 970228 following the gamma-ray emission \citep{Costa97}.  Dubbed the GRB \emph{afterglow}, similar emission was detected in several subsequent GRBs.  Optical components to the afterglow were soon observed \citep{Groot97}.  Spectra from these optical components allow redshifts to be determined, which place the progenitors of GRBs at cosmological distances from Earth \citep{Metzger97}.  

The short timescale of a GRB indicates the radiation most likely is emitted by a compact object, such as a collapsing stellar core or merging neutron star binary.  If the progenitor were to radiate isotropically, it would have to radiate away a significant amount of its rest-mass energy to be visible from Earth.  Given the required radiative efficiency of such a process, this seems extremely unlikely.  Rather the radiation is most likely emitted in a collimated fashion; as a jet directed towards Earth.  The afterglow arises when the jet begins to propagate into the medium surrounding the burst (the circumburst medium).  The subsequent expansion into the cold surrounding gas creates a relativistic blast wave propagating out from the progenitor.  Electrons in the gas emit synchrotron radiation as they are accelerated by the magnetic fields in the forward shock.  Relativistic beaming collimates the radiation in the direction of jet propagation, which is observed on Earth as the afterglow.  The radiation is dominated by x-rays in the first minutes and hours of the afterglow, but as the blast wave expands and cools over the course of days and months the afterglow progresses through the UV, optical, infrared, and radio emission.  This basic afterglow model was first outlined in the fireball model \citep{Rees92}, but in fact applies to many models.  Although the precise mechanics of GRB engines are still uncertain, the relatively simpler physics of afterglows opens them up to direct simulation and analysis.

The jet-like nature of the blast wave leads to some immediate geometric conclusions.  At early emission times, only a small patch on the leading edge of the blast wave is visible.  As time goes on, more and more of the blast wave becomes visible to the observer.  This effect slows the temporal decay of the afterglow.  However, after a long enough time, the entire blast wave becomes visible, and the afterglow will decay faster.  Further steepening of the afterglow light curve is caused by hydrodynamic spreading of the blast wave as it decelerates.  The steepening due to these processes is called the \emph{jet-break}, and is a ubiquitous effect of collimated emission.  Current observations of afterglow light curves see far fewer jet-breaks than expected, this is known as the missing jet-break problem \citep{Liang08, Racusin09}.

The jet orientations will be randomly distributed, for cosmologically distant sources such as GRBs. It has been shown that observing an afterglow from an observer angle comparable to the jet opening angle, as opposed to straight-on, can smear out the jet-break and delay the full transition until after Swift ceases observations \citep{vanEer10offaxis, vanEer11}. This provides a plausible resolution to the missing jet-break problem \citep[e.g.][]{Kocevski08, Racusin09}.  A sufficiently detailed model capable of predicting distinct light curves for off-axis observers should, in principle, allow the observer angle $\thobs$, the direction of the outflow relative to earth, to be measured from afterglow data. With enough data, this information can be used to revise predictions for the rate of jet-breaks observed by Swift (or other instruments), by revealing the extent to which observational biases will alter the intrinsic random distribution of source directions.

The hydrodynamics of a collimated outflow are inherently multi dimensional.  Calculating an afterglow light curve from basic physical parameters of the blast (such as the explosion energy, opening angle, and observer angle) requires numerical simulation since there is no known exact solution to the two-dimensional blast wave.  However, since state-of-the-art numerical simulations typically take days to run, using them directly in the analysis of data usually requires either approximations or significant amounts of computer time.  The \boxfit{} package addressed this problem with a two-fold approach \citep{vanEer12boxfit}.  Using the scale-invariance between blast energies and circumburst medium densities, fully time dependent hydrodynamical data can be generated for arbitrary values of these parameters from only a single hydrodynamics simulation.  Strong compression of the simulation output allow for data from a large sample of opening angles to be loaded into memory simultaneously.  \boxfit{} is able to generate light curves on the fly by only running a radiative transfer code using the compressed hydrodynamics data, reducing the generation time from days to seconds.  The \scalefit{} analysis code \citep[][in prep]{vanEer14scalefit} used in this work goes a step further, making use of scaling relations in the radiation equations to generate light curves directly from a precomputed table of spectral parameters \citep{vanEer12scale}.  Using \scalefit{}, an afterglow light curve can be generated directly from high resolution simulations in milliseconds. 

We make use of the speed up to perform a Markov chain Monte Carlo (MCMC) curve fitting procedure on a large sample of afterglows observed by \swift{}.  The procedure to fit a single burst requires about ten million light curves to be generated with distinct parameters, and takes a couple of hours on a standard workstation.  Any detailed model with multiple parameters is likely to exhibit degeneracies between parameters when performing a fit in only a single band.  Degenerate parameters exhibit high correlations with each other but are individually unconstrained by the data.  An advantage of the Bayesian MCMC approach is the ability to treat degeneracies as \emph{nuisance parameters}.  Nuisance parameters may be marginalized over, incorporating their uncertainties into a probability distribution for only the parameters of interest.  To efficiently sample the parameter space we use the parallel tempered affine invariant ensemble sampler \citep{Goodman10} implemented by the \emcee{} package \citep{emcee}.  While our final results leave several model parameters unconstrained, this uncertainty is folded into the estimates for the parameters we can constrain well: the jet half opening angle $\thO$, the off-axis observer angle $\thobs$, and the electron spectral index $p$.  Details of the \scalefit{} model, its implementation, and a public release of the code will be given in a forthcoming paper \citep{vanEer14scalefit}.  This paper focuses on the results of using \scalefit{} on the \swiftXRT{} dataset.  

In section 2 we give an overview of the \scalefit{} afterglow model and the simulations upon which its based.  Section 3 discusses the \swift{} data sample, and section 4 details the specific analysis we perform.  We find bursts exhibit a wide range of opening angles, but most commonly have a half-opening angle $\thO < 0.1$ rad.  The distribution of electron spectral index $p$ favours $p\sim2.2$.  The off-axis observer angle tends to be around 0.6 of the half opening angle.  The results are summarized in section 5 and discussed in section 6, with specific fit results given in the appendix.  A subset of these results were presented in \cite{Ryan13}.


\section{The Model}

The \scalefit{} GRB afterglow model uses a series of high resolution hydrodynamic simulations to calculate the time evolution of the afterglow spectral parameters.  From these parameters we extract a set of scale-invariant characteristic quantities.  The characteristic quantities depend only on $\thO$, the opening angle of the jet producing the afterglow, $\thobs$, the angle at which the observer is off-axis (the observer angle), and observer time.  Furthermore, the full set of spectral parameters depend on the characteristic quantities only through simple scaling laws \citep{vanEer12scale}.  Given the time evolution of the characteristic quantities computed from high resolution hydrodynamics simulations, this allows one to calculate the light curve for an arbitrary GRB afterglow almost instantaneously.

We employ a synchrotron model for the afterglow radiation and model the spectrum as a series of connected power laws with spectral index $p$ \citep{Sari98}:
\begin{align} \label{eq:spectrum}
	F_{fast}(\nu) &= \Fp \left \{ \begin{matrix}  \left(\nu / \nu_c\right)^\frac{1}{3} & \nu < \nu_c < \nu_m \\
								 \left(\nu / \nu_c\right)^{-\frac{1}{2}} & \nu_c < \nu < \nu_m \\
								 \left(\nu_m / \nu_c\right)^{-\frac{1}{2}} \left( \nu / \nu_m \right)^{-\frac{p}{2}} & \nu_c < \nu_m < \nu  \end{matrix}  \right . \ ,  \\
	F_{slow}(\nu) &= \Fp \left \{ \begin{matrix}  \left(\nu / \nu_m\right)^\frac{1}{3} & \nu < \nu_m < \nu_c \\
								\left(\nu / \nu_m\right)^{\frac{1-p}{2}} & \nu_m < \nu < \nu_c \\
								\left(\nu_c / \nu_m\right)^{\frac{1-p}{2}} \left( \nu / \nu_c \right)^{-\frac{p}{2}} & \nu_m < \nu_c < \nu \end{matrix}  \right . \ .
\end{align}
$F_{fast}$ ($F_{slow}$) refers to the fast-cooling (slow-cooling) regime where $\nu_c < \nu_m$ ($\nu_m < \nu_c$).  In this work we disregard self-absorption as the characteristic frequency $\nu_a$ lies well below the x-ray band observed by \swift{}.  Each of the parameters $\Fp$, $\nu_m$, and $\nu_c$ will vary with time and observer location.  The observer is located at an angle $\thobs$ off-axis at a luminosity distance $d_L$ and redshift $z$.  Furthermore the dynamics of the synchrotron radiation are parameterized by the fraction of thermal energy in electrons $\epse$, the fraction of the thermal energy in the magnetic field $\epsB$, and the fraction of electrons accelerated by the shock $\xN$.  The dependence of the synchrotron spectrum on these parameters is given by simple scaling relations \citep{vanEer12scale}.  \scalefit{} currently assumes a homogenous circumburst medium and a global cooling time (extensions are under development).  We first rescale $\Eiso$, $n_0$, and observer time $\tobs$ as:
\be \label{eq:rescale}
	\kappa \equiv \frac{\Eiso}{10^{53} \text{erg}} \ ,  \quad \lambda \equiv \frac{n_0}{1 \text{cm}^{-3}} \ , \quad \tau \equiv \left( \frac{\lambda}{\kappa}\right)^{1/3} \frac{t_{obs}}{1+z} \ .  
\ee
Then the scaling relations are given by:
\begin{align} \label{eq:scaling}
	\Fp =& \frac{1+z}{d_L^2}\frac{p-1}{3p-1} \kappa \  \lambda^{1/2} \epsB^{1/2} \xN \ \cfp(\tau; \thO, \thobs) \ , \nonumber \\
	\nu_m =& \frac{1}{1+z}\left(\frac{p-2}{p-1}\right)^2   \lambda^{1/2} \ \epse^2 \ \epsB^{1/2} \xN^{-2} \ \cfm(\tau; \thO, \thobs) \ , \nonumber \\
	\nu_c =& \frac{1}{1+z} \kappa^{-2/3}  \lambda^{-5/6} \epsB^{-3/2} \ \cfc(\tau; \thO, \thobs) \ .  
\end{align}
All the dynamic behaviour of the light curve $F_\nu (\tobs)$ is enclosed in the characteristic quantities $\cfp$, $\cfm$, and $\cfc$, which only depend on $\thO$, $\thobs$, and $\tau$.  Unfortunately $\cfp$, $\cfm$, and $\cfc$ do not have simple closed form expressions.  However, being only functions of time and two other parameters, they can easily be tabulated from simulations.  Given these tables, we have a fully physical model for all possible afterglow light curves in an ISM environment.  Without the scaling relations \eqref{eq:scaling} this would be a Herculean task, as the space of all possible light curves is (in this model) ten dimensional and would be impossible to sample at any meaningful resolution.

The current version of \scalefit{} uses the \boxfit{} simulations to calculate $\cfp$, $\cfm$, and $\cfc$ \citep{vanEer12boxfit}.  These are a series of 19 two-dimensional relativistic hydrodynamic simulations performed using adaptive mesh refinement (AMR) with the \ramcode{} code \citep{Zhang06, Zhang09}.  Each simulation calculates the time evolution of an axisymmetric relativistic jet with a particular $\thO \in [0.045, 0.5]$ rad.  The initial condition is taken to be a blast wave with a Blandford-McKee (BM) radial profile \citep{Blandford76}.  The circumburst medium has a uniform density $n_0$, and the BM solution is truncated at angle $\thO$ to provide the conical shape of the outflow.  This is consistent with the notion that at early times the outflow is ultra-relativistic and essentially radial.  As a consequence of beginning with a BM solution, our temporal coverage of $\cfp$, $\cfm$, and $\cfc$ only begins at the deceleration phase, after energy injection, plateaus, and (most) flaring has completed.  This affects what ranges of data we can reasonably expect to fit, and is discussed further in Section 3.

For typical values of $\epse$, $\epsB$, and $\xN$ the radiation does not significantly affect the dynamics of the blast wave.  Following this assumption we calculate light curves from a particular blast wave by post-processing the simulation results through a linear radiative transfer code \blastcode{} \citep{vanEer09, vanEer10transrel}. We interpolate simulation snapshots from the 19 values of $\thO$ to create snapshots for a total of one hundred $\thO$ values in $[0.045, 0.5]$ rad.  Each of these gets processed through \blastcode{} to produce a light curve $F_\nu(\tau)$ for one hundred observer angles $\thobs$.  From each light curve we extract the values of $\cfp$, $\cfm$, and $\cfc$ at one hundred values of $\tau$.  The result is three $100\!\times\!100\!\times\!100$ double-precision tables which can be used to construct arbitrary post-plateau afterglow light curves.

To construct a light curve given a set of parameters, \scalefit{} simply looks up the time series for $\cfp$, $\cfm$, and $\cfc$ at the appropriate $\thO$ and $\thobs$, applies the scaling relations \eqref{eq:scaling} to produce time series for $\Fp$, $\nu_m$, and $\nu_c$, and calculates the corresponding flux using the spectrum \eqref{eq:spectrum}.  We allow the times series for $\cfp$, $\cfm$, and $\cfc$ to be extrapolated at most one order of magnitude in $\tau$ outside the calculated tables.  Extrapolation is performed by maintaining the power-law slope of the last two entries within the tables.  The entire process takes about a millisecond to complete, allowing GRB afterglow fitting algorithms run on a laptop to generate light curves from high resolution simulations.



\section{The Data}

The \swift{} observatory was launched in 2004 \citep{SWIFT}, began taking data almost immediately and continues as of this writing.  This study includes \swift{} afterglows detected from 2005 to 2012 inclusive.  We obtained all light curve data from the \swift{}-XRT GRB light curve repository \citep{SWIFTonline}.  To reduce the dimensionality of the fit, we restrict our analysis to the 207 afterglows from this timespan with a redshift reported in the \swift{}-XRT GRB Catalogue.  

We use the count-rate light curve for our analysis.  This light curve consists of time-binned photon counts from the \swift{} X-Ray Telescope (XRT) with associated uncertainties and is automatically generated for the repository.  To convert the count rates to fluxes we employ the Counts-to-flux (unabs) conversion factors supplied in the catalogue entry for each afterglow.  These factors are the product of an automatic analysis, which uses the spectrum of observed photons and models for detector response and galactic extinction to infer the intrinsic flux associated with a single photon count \citep{SWIFTauto}.  We assume the uncertainty in this factor is small compared with the uncertainty in the count rate, and so calculate both the intrinsic flux and its uncertainty as the product of the count rate values with the Counts-to-flux factor.  

Our afterglow model does not include energy injection, pre-deceleration stage ejecta, or other effects which show themselves in the early time light curve.  By beginning the simulations with an impulsive energy injection Blandford-Mckee profile, we are deliberately modelling the afterglow only after such effects have died out.  As such, we attempt to identify and remove these features from the \swift{} data, to ensure we only perform the analysis in the regime where the model applies.  We note that if either $\nu_m$ or $\nu_c$ lies below the frequency of interest, which is true for the majority of physically relevant parameter space in the $0.3-10.0$ keV band of \swift{}'s XRT, the shallowest temporal power law slope of our model is $-1/4$ \citep{Sari98}.  This slope decays monotonically with time.  
 
Since our model cannot produce light curves with plateaus or slopes shallower than $-1/4$, we remove parts of the light curve with this behaviour as follows.  We use the best-fit broken power law provided by the \swift{} catalogue to identify the temporal phases of evolution for each afterglow.  We then cut all data points before the light curve begins a monotonic decay with a slope less than $-1/4$.  Flares identified in the catalogue entry are also cut out.  

It is possible after these cuts some light curves may have an insufficient number of data points to perform a seven-dimensional fit.  Our last requirement on the data is that a light curve have at least 3 degrees of freedom (ie. 11 data points in a 7 parameter fit) to attempt the analysis.  After all cuts were applied 188 light curves had a sufficient number of data points, on average 109 per burst.  These bursts form the sample for our study.


\section{The Analysis}

Our afterglow model has ten parameters, which we refer to collectively as $\Theta$:
\be
	\Theta \equiv \left \{  z, d_L, \Eiso, n_0, \thO, \thobs, p, \epse, \epsB, \xN \right \} \ .
\ee
We employ a Markov-Chain Monte Carlo (MCMC) approach to fit each light curve $D$ in our sample to the model.  The MCMC algorithm generates samples of the posterior probability distribution $p(\Theta | D)$, the probability distribution of the parameters $\Theta$ given the light curve data $D$.  This approach gives more information about the global structure of the posterior than a standard $\chi^2$ minimization, allowing for the identification of covariances and multimodal behaviour.  Ultimately, this method has the potential to give very accurate information about the uncertainties in the inferred values of any parameter or set of parameters.

Generating light curves from scaling laws can create degeneracies between parameters if the data lies entirely in a single spectral regime (eg. fast cooling with $\nu > \nu_m$).  Additionally, a degeneracy exists between $\epse$, $\epsB$, and $\xN$ \citep{Eichler05}.  The degenerate parameters may vary over several orders of magnitude but will produce identical light curves if certain ratios between the parameters remain fixed.  Degenerate parameters will be left unconstrained in the final analysis.  The uncertainty in these parameters will be incorporated into our estimates for non-degenerate parameters through marginalization: integrating the posterior over these \emph{nuisance parameters}.

The posterior $p(\Theta | D) $ is calculated from the likelihood $p(D | \Theta)$ and the prior $p(\Theta)$ via Bayes' Theorem:
\be \label{eq:bayes}
	p(\Theta | D) \propto p(\Theta) p(D | \Theta) \ .
\ee
The proportionality constant in \eqref{eq:bayes} accounts for normalization, and is irrelevant in the MCMC analysis.  Assuming the data points ($t_i,F_i$) are mutually independent with gaussian uncertainties $\si_i$, we can write the likelihood as a standard $\chi^2$:
\begin{align} \label{eq:likelihood}
	p(D|\Theta) &\propto \exp\left(-\frac{1}{2} \chi^2 \right) \ ,\nonumber \\
    \chi^2 &= \sum_{i} \left(\frac{F_i - F_{model}(t_i; \Theta)}{\si_i}\right)^2 \ , 
\end{align}
where $F_{model}(t_i, \Theta)$ is the flux calculated from our model at time $t_i$ with parameters $\Theta$, integrated over the \swiftXRT{} spectral band $0.3 - 10.0$ keV.  The assumption of gaussian uncertainties is valid for sufficiently high photon counts per flux measurement $F_i$.  The \swift{} data bins 15 photon counts into each $F_i$, we take this to be sufficient for Gaussian uncertainties.  

The prior $p(\Theta)$ is used to constrain and fix the parameters of the fit using prior knowledge of the data and model.  To reduce the dimension of the fit we fix $z$, $d_L$, and $\xN$.   The redshift $z$ is fixed to the value reported by the \swift{} collaboration.  The luminosity distance $d_L$ is calculated from $z$ using a benchmark $\Lambda CDM$ cosmology with $H_0 = 71\text{km}\ \text{s}^{-1} \text{Mpc}^{-1}$ and $\Omega_m = 0.27$.  There is a well known degeneracy in our formulation between $\epse$, $\epsB$, and $\xN$, to resolve this we fix $\xN = 1$ for all afterglows.

Some transformations are performed on the free parameters in $\Theta$ to improve performance.  The dimensionfull parameters $\Eiso$ and $n_0$ are made dimensionless via \eqref{eq:rescale}, parameters which may vary over several orders of magnitude are put on a log-scale, and $\thobs$ is measured as a fraction of $\thO$ (conforming to the organization of the tables).  These transformed parameters, $\Theta_{fit}$, are directly used in the MCMC routine.
\begin{align}
    \Theta_{fit} \equiv& \left \{  \log_{10} \kappa, \ \log_{10} \lambda, \ \thO, \ \thobs  / \thO, \right . \nonumber \\
    &\qquad \left . p, \ \log_{10} \epse, \ \log_{10} \epsB \right \} \ .
\end{align}
The prior on each of these parameters is uniform within certain bounds given in Table \ref{tb:bounds}.  The bounds on $\kappa$ ($\Eiso$), $n_0$ ($\lambda$), $\epse$, and $\epsB$ were chosen to contain the phenomenologically interesting regions of parameter space while eliminating unphysical regions.  The bounds on $\thO$ reflect that our numerical model is based entirely on the simulations presented in \cite{vanEer12boxfit}, which were only performed for $\thO < 0.5$. The full release of \scalefit{} will include opening angles up to $\pi/2$.  The upper bound $\thobs/\thO<1$ encodes that we must lie within the cone of the jet to observe the prompt emission.  The lower bound on $p$ is mathematically necessary for this parameterization, as $p<2$ would require an additional cut-off parameter on the accelerated particle distribution to prevent the total energy from being divergent \citep{vanEer13review}. The upper bound was chosen to be high enough to contain the physically likely values.

In code tests (see Section 4), we found fits were not very sensitive to the overall scale of the light curve.  Raising or lowering all the flux values by 20\% did not induce significant changes in our conclusions.  This is a benefit, as it means our analysis are somewhat insensitive to the exact value of Counts-to-flux used, particularly to the specifics of accounting for dust absorption and galactic extinction. 

\begin{deluxetable}{cc}
\tablecaption{Bounds on uniform prior for parameters in $\Theta$. \label{tb:bounds}}
\tablewidth{0.3\textwidth}
\tablehead{\colhead{Parameter} & \colhead{Bound}}
\startdata
$\log_{10} \kappa$ & $[-10.0,3.0]$  \\[2pt]
$\log_{10} \lambda$ & $[-5.0,5.0]$ \\[2pt]
$\thO $& $[0.045,0.5]$ \\[2pt]
 $\thobs / \thO$ & $[0.0,1.0]$ \\[2pt]
$p$ & $[2.0, 5.0]$ \\[2pt]
 $\log_{10} \epse$ & [-10.0, 0.0] \\[2pt]
$\log_{10} \epsB$ & [-10.0, 0.0] \\
\enddata
\end{deluxetable}

The MCMC analysis is performed using the parallel-tempered affine-invariant ensemble sampler implemented by the \emcee{} python package \citep{Goodman10, emcee}.  This algorithm uses an ensemble of ``walkers'' moving simultaneously through parameter space instead of the standard single walker approach (e.g. Metropolis-Hastings).  The use of the ensemble allows the method to be affine-invariant: affine transformations of the distribution do not affect the efficiency of the sampling.  Strong linear correlations between parameters are sampled with the same quality as uncorrelated parameters, a difficult problem for traditional samplers.  

Parallel tempering is a technique used to better sample multimodal distributions (\citet{Swendsen86, Geyer91}, see \citet{Earl05} for a review).  Several ensembles are run simultaneously on likelihoods $p(D|\Theta)^{1/T}$, where $T$ is the temperature.  The lowest temperature ensemble $T=1$ samples the true posterior, while hotter ensembles sample distributions which more closely resemble the prior and hence are less restricted in their movements.  The ensembles are coupled by allowing walkers to swap between them with some probability every iteration.  This allows walkers to mix between distant modes, allowing efficient sampling of multimodal distributions.  Parallel tempering is tuned by the choice of temperature ladder: the temperature of each ensemble.  We use the default ladder provided with \emcee{}: a geometric sequence of user set length and growth factor tuned to optimally sample an n-dimensional gaussian.  Only samples from the true distribution ($T=1$) are used in the analysis.

The \swift{} XRT band lies above $\nu_m$ and $\nu_c$ in a large region of parameter space, so a large number of light curves generated by \scalefit{} will exist in only a single spectral regime.  In this case the scaling relations used to calculate the flux will create a large degeneracy between $\Eiso$, $n_0$, $\epse$, and $\epsB$, leading to high correlations between these parameters.  Affine-invariance is thus a highly beneficial property of the sampler for this analysis.

Each afterglow was sampled with 20 temperature levels and 100 walkers per level, centered in a small ball around the reference point in phase space:
\be \label{eq:init}
	\Theta_{fit, 0} = \{0.0, \ 0.0,\  0.2, \ 0.5,\  2.5, -1.0, -2.0\} \ .
\ee
Walkers were initialized with uniform random values within 2\% of \eqref{eq:init} in each non-zero reference value, and within $\pm 0.02$ if the reference value was zero.  There currently does not exist any unbiased method for determining the convergence of an MCMC chain.  As such we employed a pragmatic view for determining the ``burn-in'' and running lengths of the chain.  Initial tests indicated the autocorrelation time of the chain to be approximately 1000 iterations, depending on the parameter.  Based on this we chose a burn-in of 7000 iterations, to ensure several autocorrelation times between the initial condition and the recording of samples.  Sampling was performed for 3000 iterations after burn-in, generating altogether 300 000 samples of $p(\Theta_{fit} | D)$ for each burst.

For each burst the final values of the fit parameters are inferred from their marginalized posteriors (eg. $p(\thobs |D)$) estimated by the MCMC generated samples.  The marginalization incorporates the uncertainties in all other fit parameters.  Our estimate for each parameter is the median of the posterior, with uncertainty given by the $68\%$ quantiles.  We use the median (instead of the mean or mode for instance) because it is less sensitive to the tails of the distribution and it is preserved under invertible mappings of the parameter (eg. $\log_{10} \epse \to \epse)$.

	
We validated this analysis by performing fits on a set of synthetic data.  Using the \scalefit{} model we produced a set of 500 light curves with randomly distributed parameters $\Theta$ in the \swiftXRT{} 0.3-10.0 keV band, using the method from \citet{vanEer11}.  Data points were drawn from each light curve in a manner to resemble \swift{} data.  Occlusion by the Earth was taken into account by only taking data in 48 minute intervals, and the fractional exposure of late-time afterglows was included by reducing the data rate by a factor of ten for $t_{obs} > 1$ day.  Fluxes were given random gaussian uncertainties (25\% for early time and 50 \% for late time).  Each of these synthetic light curves was subjected to our analysis using a single ensemble of 1500 walkers and no parallel tempering.    

\begin{figure*}
    \plotone{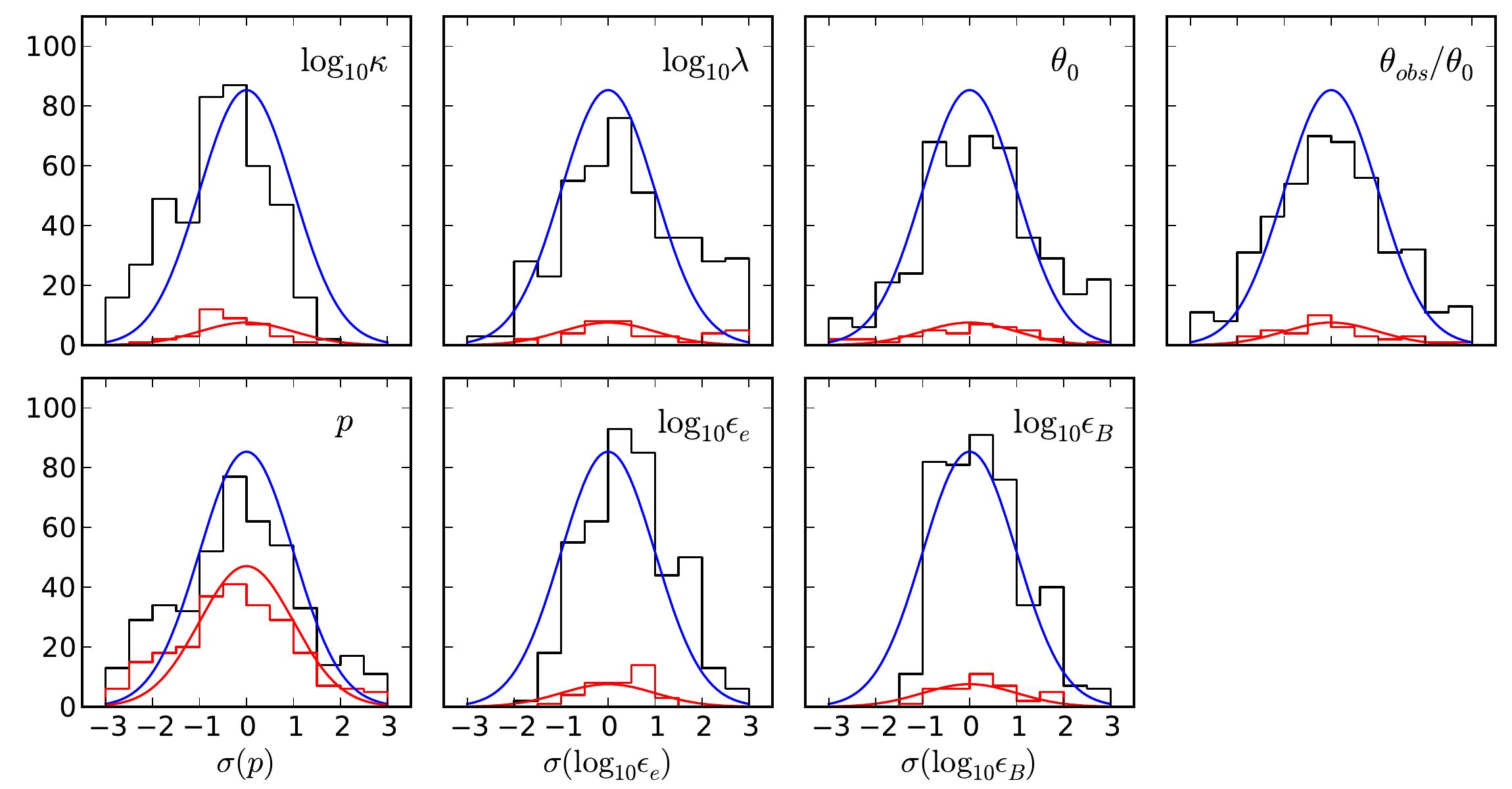}
	\caption{\label{fi:test} The discrepancy ($\si$) between true and estimated values of $\Theta_{fit}$ in fits to synthetic data are shown in the black histograms.  A discrepancy $ \si=\pm 1$ ($\pm 2$) indicates the true value lies on the 68\% (95\%) quantile of the estimate.  The blue curve shows the theoretical distribution of discrepancies: a gaussian normalized to the number of synthetic datasets with unit width.  The red curve and histogram shows the same distributions after cutting to ``well-fit'' curves (our definition of ``well-fit'' is introduced in Section 5).  These results were run with a single ensemble of 1500 walkers, no parallel tempering.}
\end{figure*}

Fig \ref{fi:test} shows the a histogram of discrepancies between the true and estimated values for $\Theta_{fit}$ for each synthetic afterglow.  Theoretically these distributions should be approximately Gaussian.  The tails of the distributions for $\kappa$, $\lambda$, and $\kappa$ indicate this analysis may underestimate $\Eiso$ ($\ka$) and overestimate $n_0$ ($\lam$) and $\epse$.  Due to the degeneracies in these parameters when fitting only a single band, this is not entirely surprising.  The distribution of discrepancies for $\epsB$, $\thO$, $\thobs/\thO$, and $p$ agree with the expected Gaussian shape, indicating the analysis is at least internally consistent on these parameters.  We believe this is sufficient indication that \scalefit{} can provide good estimates of $\thO$, $\thobs / \thO$, $p$, and $\epsB$ when fitting single band x-ray afterglow light curves.  Further runs on the synthetic data demonstrated our results are not altered when fitting the spectral flux $F_\nu$, flux $F$, or fluence $\int F dt$.

In testing it was discovered $p(\Theta_{fit}|D)$ may be strongly multi-modal, particularly when light curves are allowed to extrapolate outside the calculated tables for $\cfp$, $\cfm$, and $\cfp$.  In some cases, the multi-modality is strong enough that parallel tempering cannot adequately sample the distribution.  To minimize the effect of these cases we adopted the following heuristic.  Every light curve is fit twice, once allowing extrapolation (to a maximum of one order of magnitude in $\tau$) and once with no extrapolation.  The results we report are from the run that found the lowest $\chi^2$: the sample with highest likelihood.    

A parallel effort has been performed using the same model theoretical model with an independent implementation \citep{Zhang14}.  This study focuses on a smaller number of bursts which were observed by both \swift{} and \chandra{} and uses \multinest{} sampling for its analysis.  Both methods were tested during development for convergence, sensitivity to initial conditions, and behaviour with resolution.  The results were consistent between both analyses, demonstrating the robustness of the methods and providing a measure of independent validation.


\section{The Results}

We ran the \scalefit{} analysis on all 188 afterglows in our sample.  A table of results for the non-degenerate parameters ($\thO, \thobs/\thO, p$) is given in the appendix.  The quality of the fits varied from burst to burst.  Some bursts had very sharp fits, most had a broad (ie. unconstrained) distribution in at least one of the parameters of interest.  A small minority of fits ($\lesssim10\%$) were unable to converge to an adequate light curve, with a best fit $\chi^2/dof \gg 1$.  Given the freedom in our model, we expect the afterglows in the latter category do not satisfy one or more of our base assumptions, perhaps that of a homogenous ISM.

\begin{figure*}
\centering
\begin{overpic}[width=\textwidth]{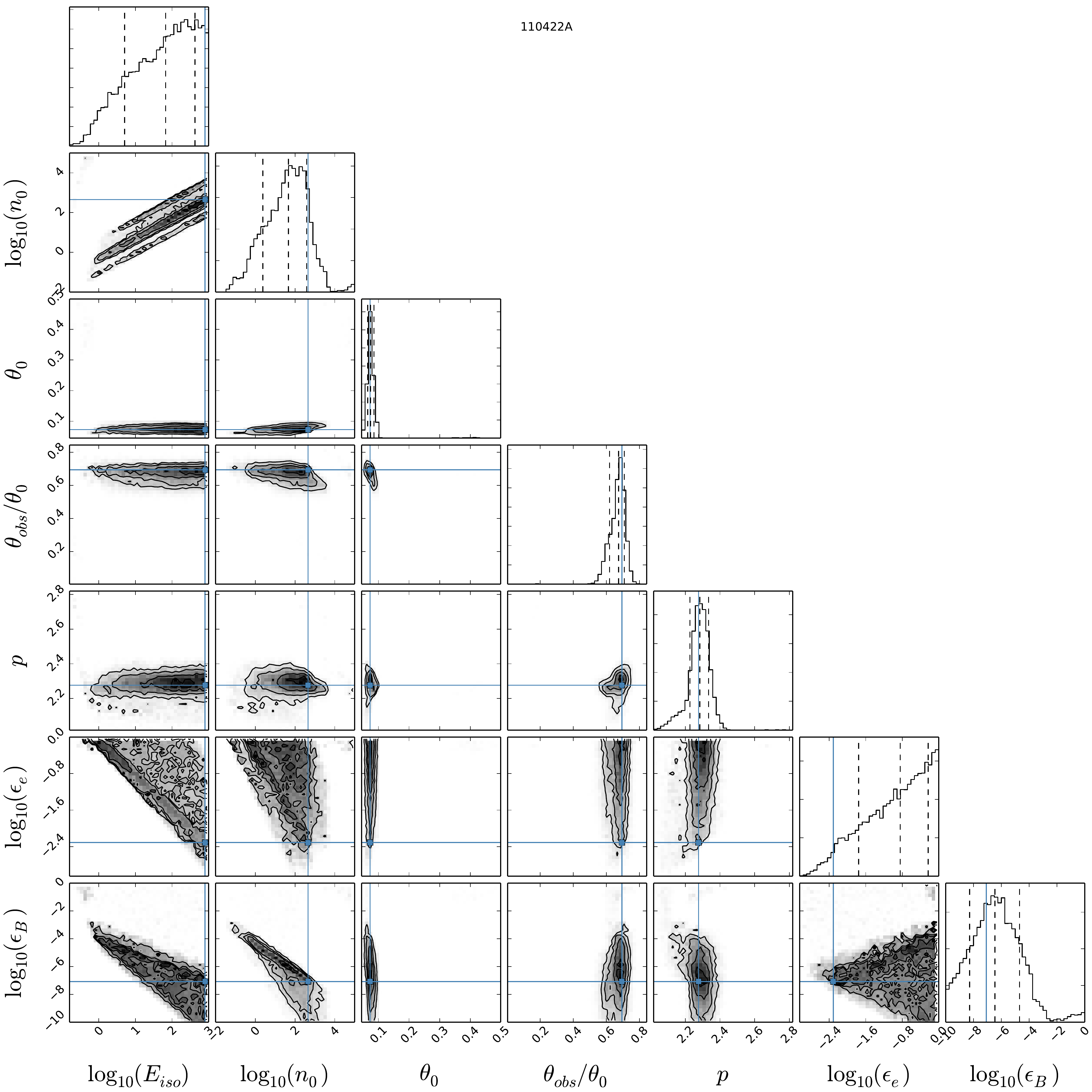}
	\put(46,60){\includegraphics[width=0.55\textwidth]{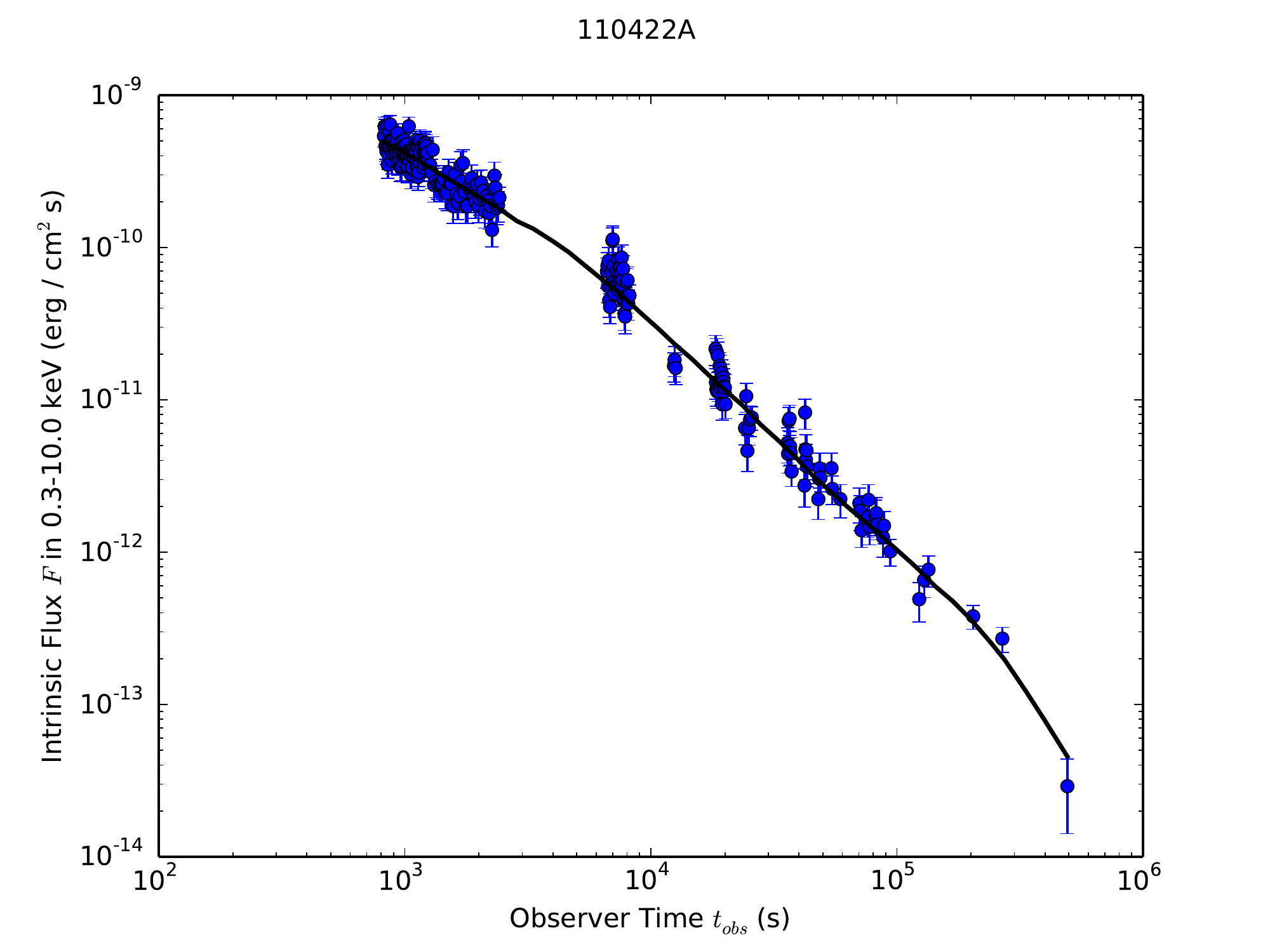}}
\end{overpic}
\caption{Fit result for 110422A.  The diagonals along the corner plot show the marginalized probabilities for each parameter.  The off-diagonal contour plots show the covariances between all pairs of parameters.  The best-fit values (MAP, maximum posterior probability) are shown in blue.  The best-fit light curve is shown against the data in the upper right.}
\label{fig:fit1}
\end{figure*}

The fit for GRB 110422A is shown in Fig \ref{fig:fit1}. It serves as an example of a particularly good fit.  The corner plot shows projections of $p(\Theta_{fit} | D)$ as determined by the MCMC samples.  Plots on the diagonal show the marginalized distributions for each individual parameter, while the off-diagonal plots show the pairwise correlations between parameters.  We see the distributions of $\ka$, $\lam$, $\epse$, and $\epsB$ are very broad, covering several orders of magnitude.  This of course is due to the model's degeneracy between these parameters, which induces a strong correlation between them. This correlation is exhibited in the off-diagonal plots for these parameters.  The energy and circumburst density ($\ka$ and $\lam$) are confined to narrow bands in phase space, while $\epse$ and $\epsB$ exhibit a multidimensional degeneracy.  This degeneracy does not affect the distributions of $\thO$, $\thobs/\thO$, or $p$, as can be seen from their plots on the diagonal.  Despite several other parameters being unconstrained, these parameters are determined quite well.  We find $\thO = 0.0733^{+0.011}_{-0.0098}$, $\thobs/\thO = 0.676^{+0.035}_{-0.050}$, and $p=2.284^{+0.049}_{-0.06}$.  In particular, this afterglow was almost certainly observed off-axis.

In general, the posterior distributions of $\Eiso$ and $n_0$ tend to be very broad or uniform, with a tight pairwise correlation between them.  The correlation is caused by the rescaling of $t_{obs}$ to $\tau$, which depends on the ratio $\Eiso/n_0$.  The radiation parameters $\epse$ and $\epsB$ tend to demonstrate a more multidimensional degeneracy with $\Eiso$, $n_0$, and each other.  These reflect the inherent degeneracy in our model when restricted to single band fits.  $\epsB$ is least constrained by most fits.  This is not unexpected as the overall dependence of the flux $F$ on $\epsB$ is very weak.  For example, above the cooling break $F_\nu \propto \epsB^{(p-2)/4} \approx \epsB^{0.05}$ using our median value of $p$.  We expect these difficulties to ease in multi-band fits.

On the other hand $\thO$, $\thobs$, and especially $p$ are constrained well by several fits.  This is not surprising, as the dependence of the light curve on these parameters is not given by simple power-law scaling.  With these estimates for $\thO$, $\thobs$, and $p$ we can make some statements about the global distribution of these parameters amongst the \swift{} sample.  

\begin{figure*}
    \plottwo{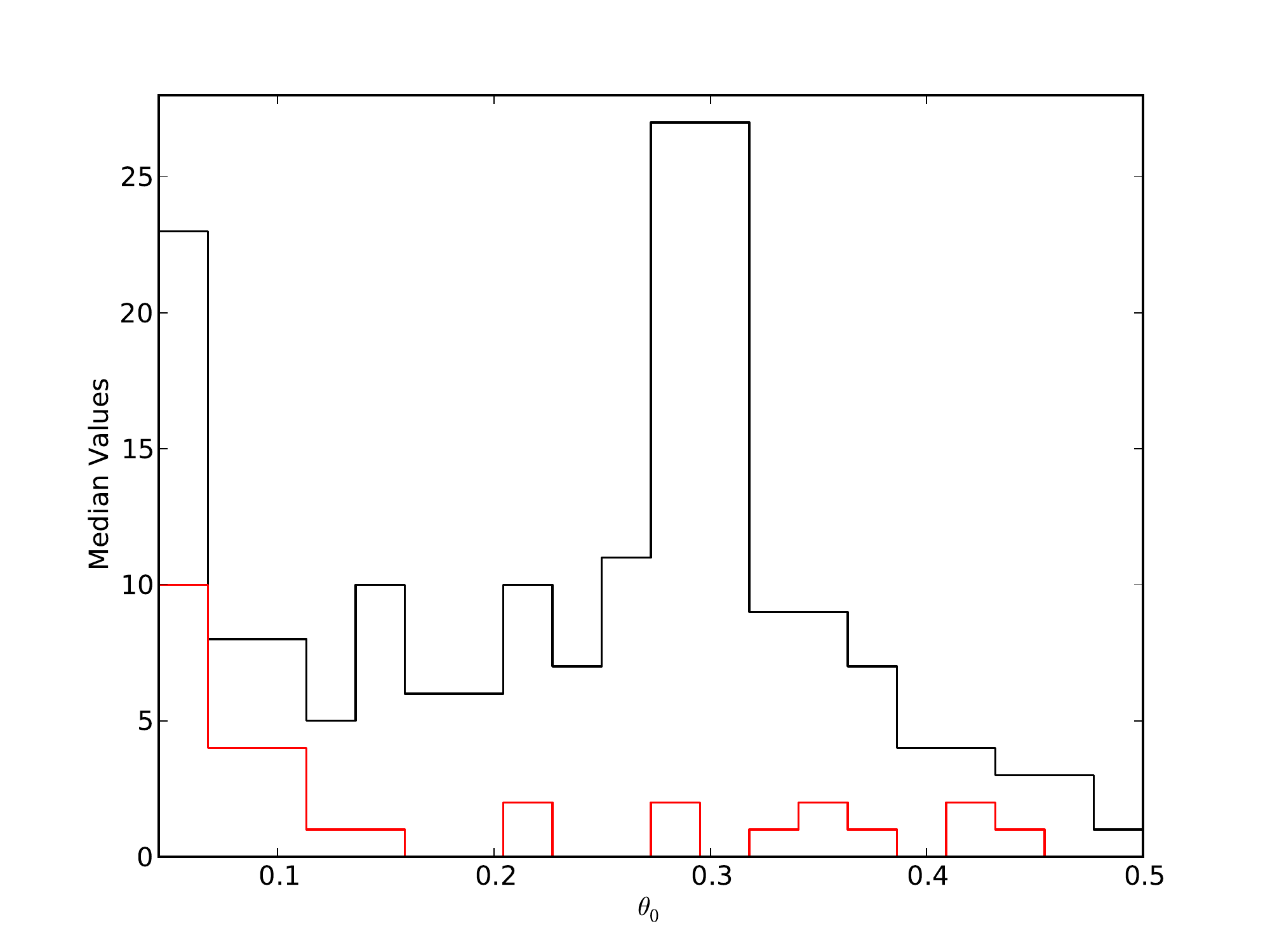}{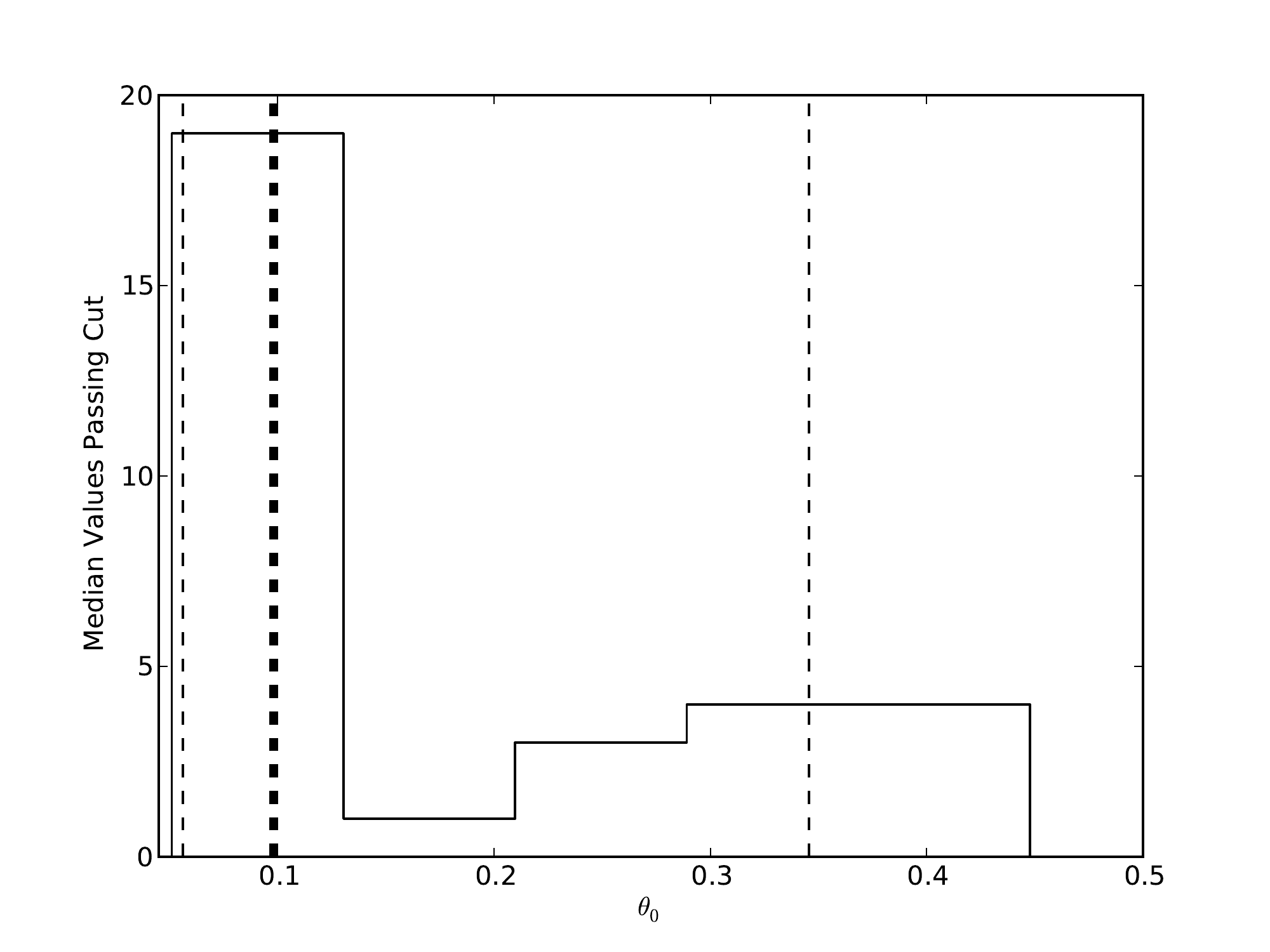}
	\caption{\label{fi:thO} Distribution of $\thO$ across all afterglows in the sample. The black curve in the first figure shows the distribution of reported values of $\thO$ (eg. the median of the marginalized distribution for each light curve).  The red curve plots the same, including only bursts considered ``well-fit'' in $\thO$ and $\thobs$.  The second figure shows only the ``well-fit'' bursts, with the median (thick dashed) and  68\% quantile (thin dashed).}
\end{figure*}

\begin{figure*}
    \plottwo{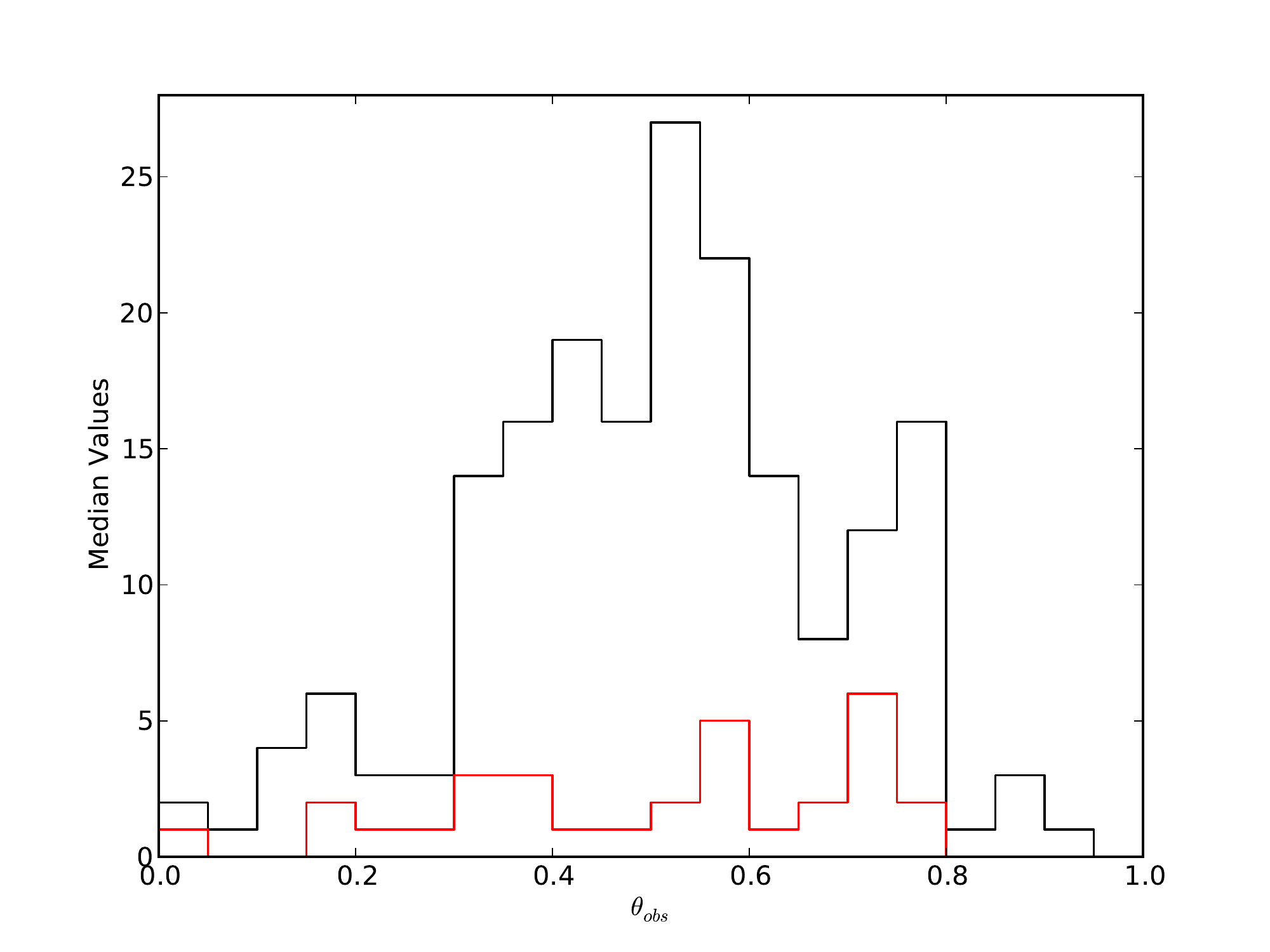}{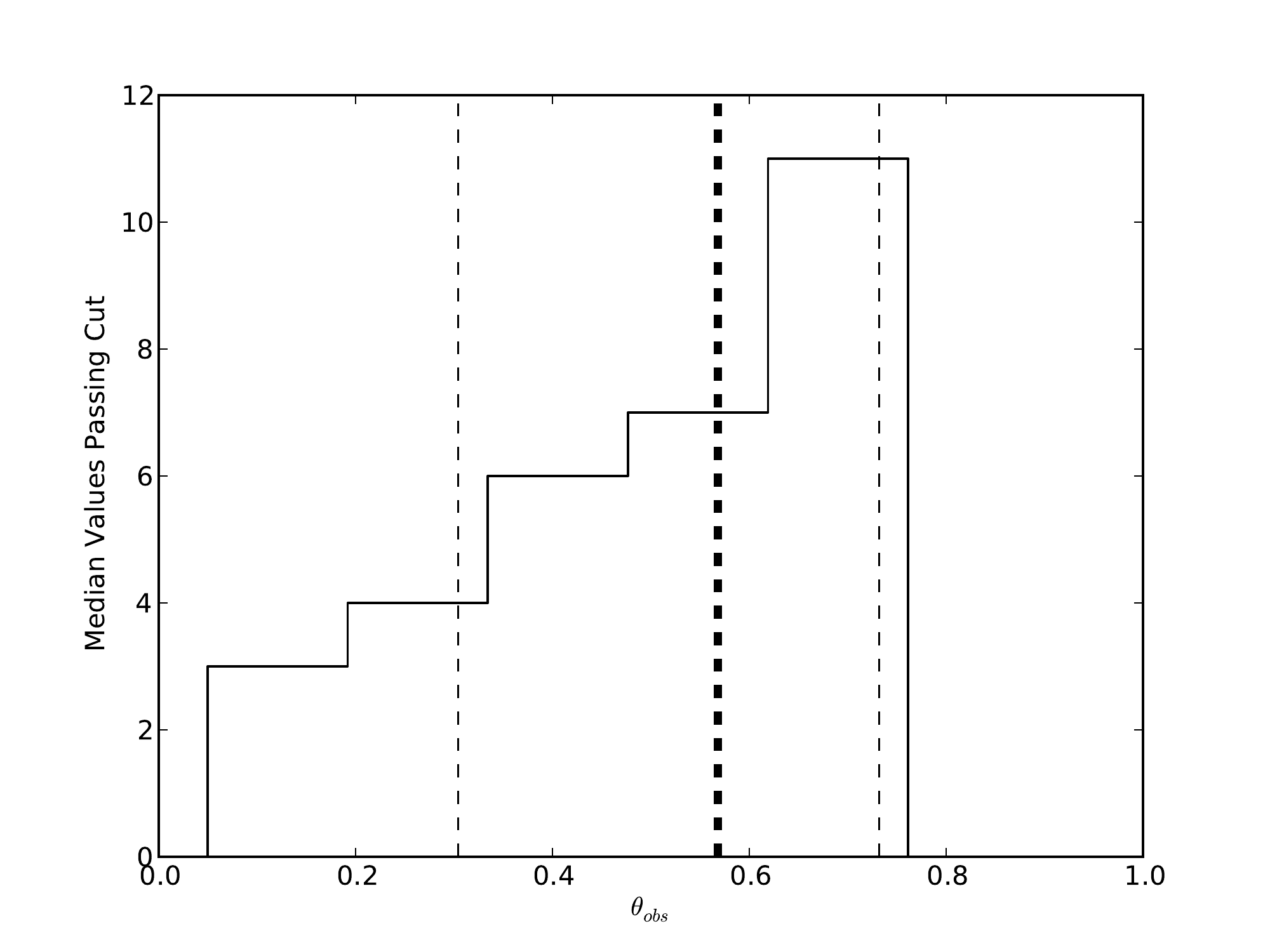}
	\caption{\label{fi:thobs} Same as Fig. \ref{fi:thO}, but for $\thobs / \thO$.}
\end{figure*}

\begin{figure*}
    \plottwo{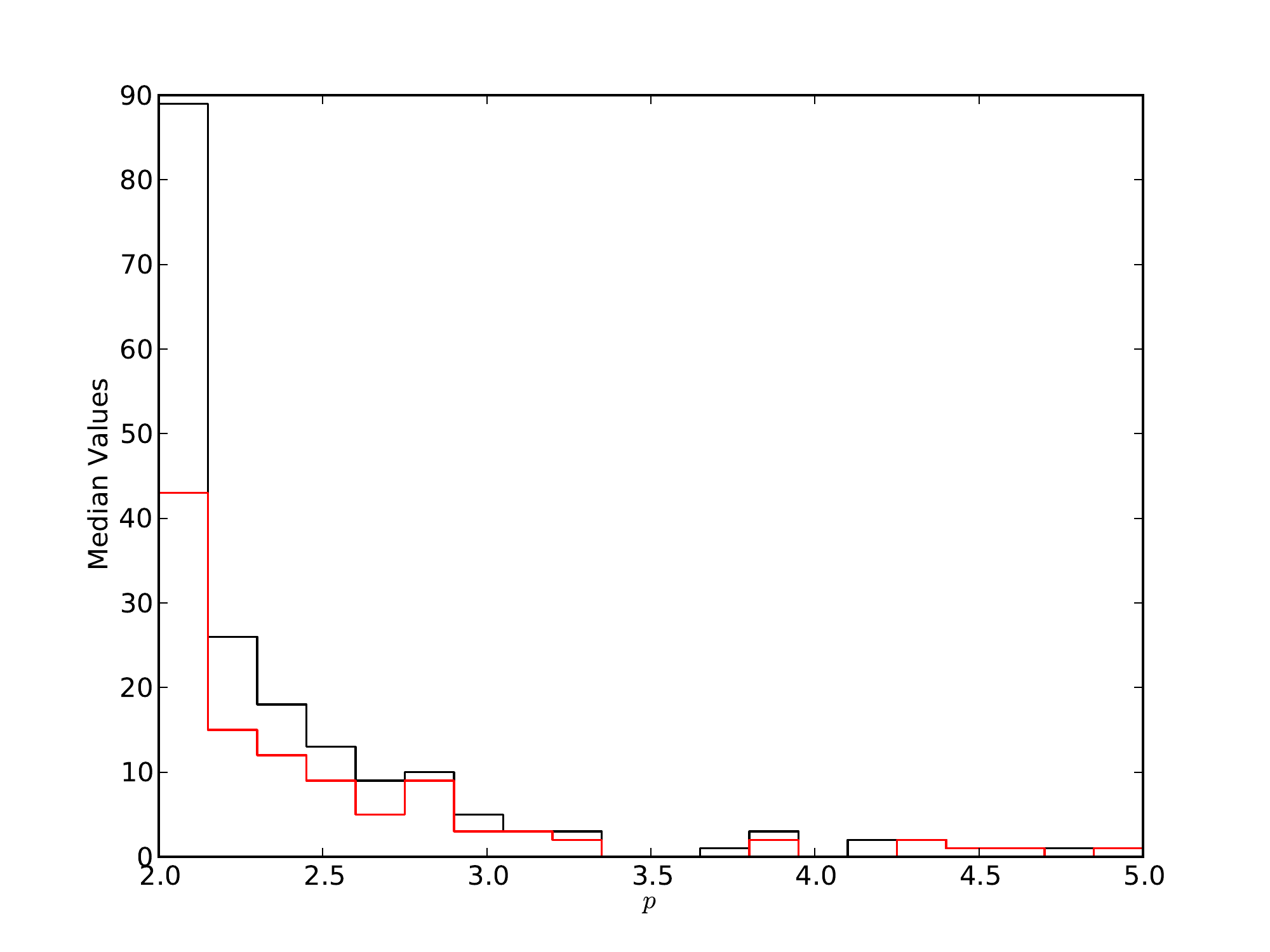}{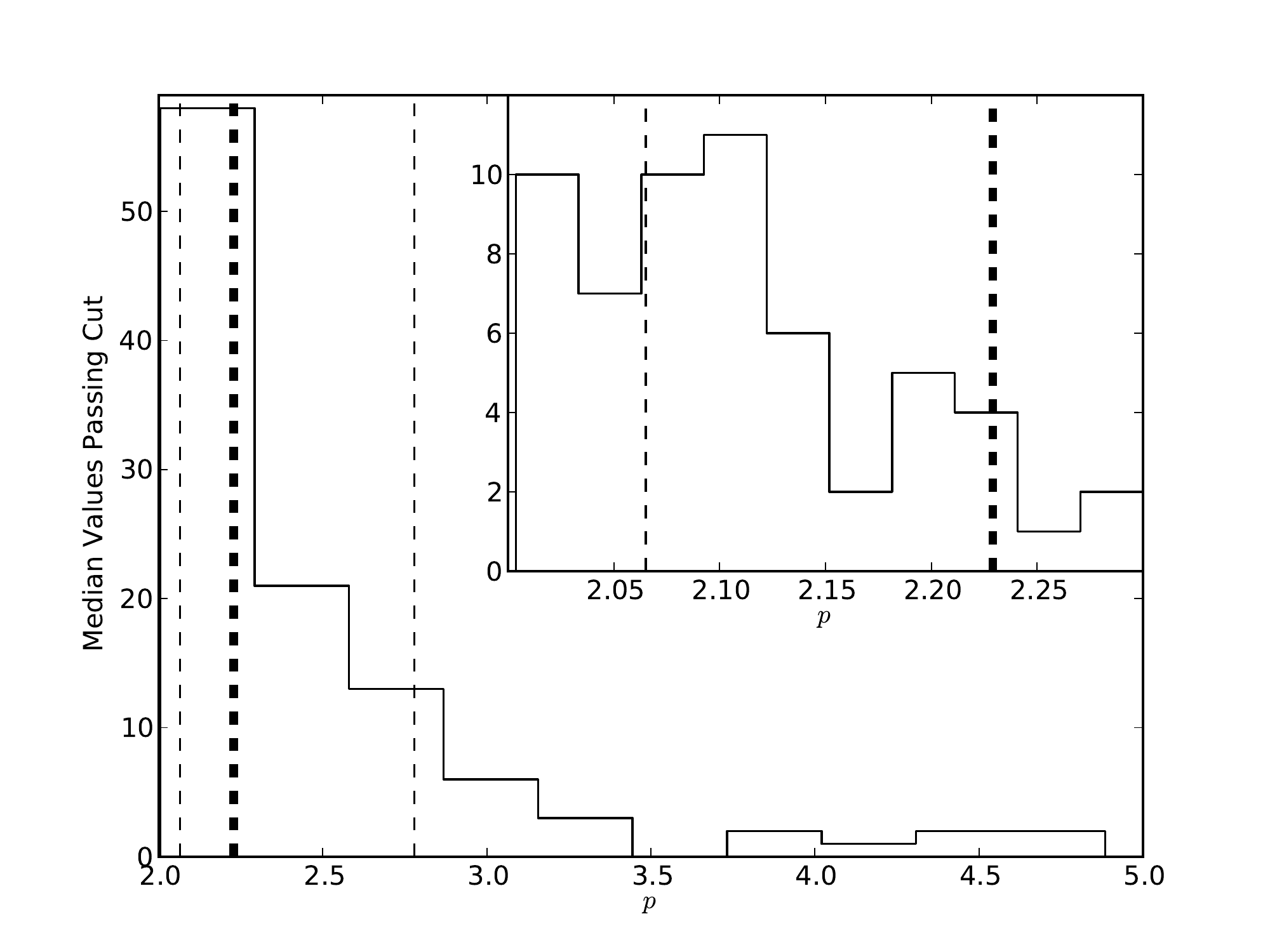}
	\caption{\label{fi:p} Same as Fig. \ref{fi:thO}, but for $p$.  The ``well-fit'' criteria is applied only to $p$ and the inset is zoomed to $2.0 < p < 2.3$.}
\end{figure*}

In first panels of Figs \ref{fi:thO}, \ref{fi:thobs}, and \ref{fi:p} we plot histograms of the estimated values for $\thO$, $\thobs/\thO$, and $p$ for our sample.  The opening angle $\thO$ has a broad distribution with isolated peaks at $\thO \sim 0.05, 0.28$.  The observer angle $\thobs / \thO$ is broad, with a spike at $\thobs/\thO \sim 0.5$. The spectral index $p$ decreases from $p\sim 2.1$ into a tail which extends to $p=3$.  Since these plots include fits which had unconstrained distributions, they do not reflect the true distributions $P(\thO)$, $P(\thobs)$, $P(p)$ in our sample.  For instance, the large spikes near the domain center in Figs. \ref{fi:thO} and \ref{fi:thobs} are due to including the medians of distributions which were almost uniform.   

To resolve this we make cuts on our dataset to eliminate poorly-constrained fits.  We employ the following criterion to qualify a particular parameter as ``well-fit'': the width of the 68\% quantile must be less than half the width of parameter's domain (see Table \ref{tb:bounds}) and the best-fit (maximum posterior, minimum $\chi^2$) value must lie within the 68\% quantile.  The number of light curves passing each criteria are given in Table \ref{tb:passed}.

\begin{deluxetable}{cc}
\tablecaption{Number of light curves passing the ``well-fit'' criteria for each parameter. \label{tb:passed} }
\tablewidth{0pt}
\tablehead{\colhead{Parameter} & \colhead{Number of Well Fit Light Curves}}
\startdata
$\thO$ & $52$  \\
$\thobs / \thO$ & $63$ \\
$p $& $108$ \\
 $\thO$ and $\thobs / \thO$ & $31$ \\
$\thO$,  $\thobs / \thO$ and $p$ & $17$ \\
\enddata
\end{deluxetable}

For a burst to be considered ``well-fit'' and included in the global distribution of $\thO$ and $\thobs$ we require it to have passed the criteria for both $\thO$ and $\thobs$.  There are 31 such light curves in our sample, the histograms of their values of $\thO$ and $\thobs / \thO$ are given in the second panels of Figs. \ref{fi:thO} and \ref{fi:thobs} respectively.  For a burst to be included in the global distribution of $p$ we only require it to be ``well-fit'' in $p$.  There are 108 of these bursts, the histogram of their values of $p$ is given in the second panel of Fig \ref{fi:p}.  Summary statistics of these histograms are given in Table \ref{tb:quantiles}.

\begin{deluxetable}{cccc}
\tablecaption{Quantiles on distribution of fit values for $\thO$, $\thobs / \thO$, and $p$. \label{tb:quantiles} }
\tablewidth{0pt}
\tablehead{\colhead{Parameter} & \colhead{Median} & \colhead{68\% Quantile} & \colhead{95\% Quantile}}
\startdata
$\thO$ &  $0.098$ & $(0.056, 0.35)$ & $ (0.055, 0.42)$\\
$\thobs / \thO$ & $0.57$ & $(0.30, 0.73)$ & $(0.18, 0.75)$ \\
$p $& $2.23$ & $(2.07, 2.78)$ & $(2.02, 3.8)$ \\
\enddata
\end{deluxetable}

\underline{Half opening angle $\thO$}: We find a full half of the well-fit bursts have a small opening angle $0.045 < \thO < 0.098$ rad.  The remainder are broadly distributed throughout the allowed range $0.098 < \thO < 0.5$ rad.  As the lower cutoff at 0.045 rad is a reflection of our simulation coverage and prior, it is possible some of the small angle bursts in fact have $\thO < 0.045$ rad.  Our results are consistent with \cite{Racusin09}, who found a similar distribution of $\thO$ with a median of 0.094 rad.

\underline{Off-axis observer angle $\thobs$}:  The $\thobs  / \thO$ distribution is broad with a median at 0.57 and 95\% quantiles at (0.18, 0.75).  These afterglows are almost certainly observed off-axis, at a significant fraction of the opening angle.  This can have a profound effect on afterglow light curve analysis.  Jet-breaks occur when the emission surface seen by an observer begins extending beyond the edges of a jet, causing the light curve to sharply steepen.  When viewed off-axis the near edge of the jet will be seen before the far edge, causing the jet-break transition to become extended over time  \citep{vanEer12obs, vanEer13boost}.  It may be very difficult to see off-axis jet breaks without late-time observations.

\underline{Electron spectral index $p$}:  More than half of our entire sample passed the well-fit criteria for $p$.  The distribution of fit values for $p$ spans the allowed domain, favouring smaller values $p < 2.23$.  Some concern may be raised that so many bursts seem to require values so close to the $p>2$ boundary imposed by the prior.  From inspection of individual fits, we find this is not the case.  Bursts with well-fit $p$ distributions centered near 2 tend to be very sharply peaked, so that the $p=2$ case is safely in the tails (eg. Fig \ref{fig:fit1}).  Similarly, we find no indication of interference from the $p<5$ upper bound.  

\subsection{Short Bursts}

Eight short GRBs were included in our sample\footnote{The short GRB 050509B did not have enough data points to be included.}.  Unfortunately few of them produced useful results from our analysis.  GRBs 060801, 070724A, 080905A, 100117A, and 101219A were unable to find a good fit: all had very broad distributions and a minimum $\chi^2/dof  > 2.0$.  GRBs 090426 and 100724A found reasonable fits, but still do not have enough data to constrain the parameters.  These bad fits could be the result of being over-aggressive in our data selection, inappropriate priors, or that these afterglows break one or more of our model assumptions.  

On the other hand, GRB 051221A produced a good fit with $\thO = 0.448^{+0.031}_{-0.038}$, $\thobs/\thO = 0.449^{+0.053}_{-0.058}$, and $p = 2.024^{+0.025}_{-0.014}$.  This indicates the afterglow was most likely caused by a very broad jet, and was observed significantly off-axis.  The $p$ value is quite small, with the distribution running directly into the lower bound at $p = 2.0$.  This indicates a need for followup with a model that can provide for $p < 2$.

\citet{Zhang14} find a different best-fit: $\thO = 0.10$, $\thobs/\thO = 0.08$, and $p=2.36$.  Their results are consistent with ours, however, as their posterior distributions are quite broad and have significant weight in our reported region.  Our fit uses only \swift{} data and begins at $t_{obs} = 1877$ s, while their fit includes late time \chandra{} data and only begins at $t_{obs} = 30\ 864$ s.  To determine the exact source of the difference we did runs with both starting times, both with and without the \chandra{} data.  The \chandra{} data does not strongly affect the fit, rather it is the choice of initial time which strongly determines the result.  

The light curve for 051221A has a plateau from $\sim$2000 s to $\sim$20,000 s which we include as part of the early time segment as it passes the protocol outlined in Section 3.  A detailed analysis of this burst using radio, optical, and x-ray data was given in \citet{Soderberg06}, reporting $0.10 \leq \thO \leq 0.13$, $p = 2.15 \pm 0.10$, $\thobs=0$ (implicitly) and attributing the plateau to an energy injection phase.  Our best-fit, ignoring self-absorption, matches both the radio and x-ray data but systemically under predicts the optical data by a factor of a few.  This discrepancy with the optical could be due to the global cooling time approximation \citep{vanEer10offaxis, Guidorzi14}.  A strong caveat on our fit result is the assumption the plateau is due solely to the observational effects of a decelerating blast wave.  In particular it is difficult to reconcile the regular pre-plateau stage with a pure blast wave model.  If the plateau is indeed due to energy injection, then the \scalefit{} model is inapplicable to the early-time light curve and a later time to begin fitting would be more appropriate.  

Constraining the orientation, $\thobs$, and opening angle $\thO$, of short GRBs remains of specific interest because of the implication for gravitational wave science. If short GRBs are detectable electro-magnetic counterparts to gravitational wave events, constraining their orientation would greatly reduce the degeneracy in the possible fits to the GW signal \citep{Nissanke13}.  Additionally, constraining their opening angle will constrain their beaming corrected rates and total energies.


\section{Discussion}

Given the usual cosmological assumptions of isotropy and homogeneity, it is expected that the orientation of GRB jets is random.  Since larger off-axis angles $\thobs$ correspond to a larger solid angle of possible viewing, the theoretical distribution of $\thobs/\thO$ for a homogeneously emitting conical outflow is linear:
\be
	P_{theo}(\thobs/\thO) \propto \thobs / \thO \ . \label{eq:theo}
\ee
However, this is almost certainly not the correct distribution of \emph{observed} $\thobs/\thO$.  The likelihood of observing an afterglow with a particular $\thobs/\thO$ depends not only on the orientation of jets relative to earth (eq. \ref{eq:theo}), but also on the spreading of the blast wave, the brightness profile across the blast wave, and observational detection biases \citep[see e.g.][]{vanEer12obs}.  The brightness profile of a blast wave has two sources: the structure of the outflow itself and the observational limb-brightening effect.  The structure of outflows arising from GRBs is still unknown.  Many models exist in the literature, including the basic unstructured top-hat and more complicated structured models such as two-component \citep[e.g.][]{Berger03}, power-law decay \citep{Rossi02}, or the boosted fireball \citep{Duffell13}.  Regardless of the outflow structure, the simple optics of observing a relativistic outflow also produce a limb-brightening effect, enhancing observed radiation from the on-edge region of a blast wave \citep{Meszaros06}.  Both these effects will be rolled into the observed distribution of $\thobs/\thO$, as well as any observational biases which may be present for the particular experiment under consideration.  In principle, given enough data it may be possible to study the structure of the outflow from the observed distribution of $\thobs/\thO$.

Our distribution of $\thobs/\thO$, shown in Fig \ref{fi:thobs}, indicates many afterglows have significant observer angles and very few have $\thobs < 0.2\ \thO$ or $\thobs > 0.8\ \thO$.  Although we are not currently able to make claims to the source of these features, be they due to orientation, structure, or observational biases, the sheer presence of large observer angles in several bursts is of crucial importance to estimates for total jet energy $E_{jet}$.  

Figs \ref{fi:thO} and \ref{fi:thobs} indicate a typical observer angle is $\thobs \approx 0.05$ rad $\approx 3$ degrees.  It is important to state that while these angles are small, they are not insignificant.  Even small off-axis observer angles can have a large effect on the light curve if the opening angle was also small.  In particular, when $\thobs$ is a significant fraction of $\thO$, the jet break is smeared from a sharp feature to an extended transition from pre- to post-break slopes.  The delay in the onset of the post-break slope can be large enough to push the jet break out of \swift{}'s typical viewing window (i.e. 10 days) \citep{vanEer10offaxis, vanEer11}.  Additionally, the smooth transition may hinder efforts to detect jet breaks by broken power-law fits.  Both these effects likely contribute to the  ``missing jet break'' problem.
  
Our results are generally consistent with the complementary study investigating afterglows with observations by both \swift{} and \chandra{} \citep{Zhang14}.  This study used the same theoretical model, but with an independent implementation and \multinest{} sampling.  The overall consistency of the results shared by both studies validates the robustness of the model and analyses.  Some individual fits display differences, most of which are due to different choices in selecting how much of the \swiftXRT{} data to attempt to fit.  The inclusion of \chandra{} points can also have a strong effect on the light curve, indicting the utility of using late time followup when it is available.

Possible methods for improving the fit results are currently being studied.  Further work will include the incorporating multi-band data and and a stellar wind circumburst medium into the fits.  Multi-band data should break the degeneracy between $\Eiso$, $n_0$, and $\epse$, allowing much more information to be extracted from the analysis.  Inclusion of stellar wind environments will hopefully lead to more well-fit bursts in this sample.

The full results of this analysis, the posterior probability distribution function $p(\Theta | D)$ of each burst in the sample, will be made available online at the Afterglow Library\footnote{\afterglowlibrary}. Researchers interested in using results of this paper are highly recommended to use these instead of the summary statistics presented in the appendix of this paper.  The full 7-dimensional pdf, for instance, may be marginalized to produce a pdf for any subset of $\Theta$, taking into account all correlations between parameters.  The posterior $p(\Theta | D)$ will be made available as a list of MCMC samples in the HDF5 data format.


\section{Summary}

We have run the \scalefit{} analysis on a sample of GRB afterglow light curves observed by \swift{} between 2005 and 2012.  The sample included all afterglows with a known redshift and a sufficient number of data points to perform the analysis; 188 light curves in total.  The \scalefit{} afterglow model uses scaling relations in the hydrodynamic and radiation equations to calculate a light curve from precomputed 2d simulations.  Although the model displayed significant degeneracies between $\Eiso$, $n_0$, $\epse$ and $\epsB$, the values of the opening angle $\thO$, observer angle $\thobs$, and spectral index $p$ could be constrained in several bursts.  The \scalefit{} package will be released to the public in the near future \citep{vanEer14scalefit}.

Of the 188 fits, 108 had sufficiently constrained values of $p$ to be included in the result.  We found $p$ has a highly asymmetric distribution, with a median of 2.23 and 68\% quantile $(2.07 , 2.78)$.  Thirty one bursts in our sample were sufficiently constrained in both $\thO$ and $\thobs / \thO$ to be included in the result.  The distribution of $\thO$ was also highly asymmetric, with a median of $0.098$ rad and a 68\% quantile $(0.056, 0.35)$.  The off-axis observer angle $\thobs/\thO$ had a median of $0.57$ with a 68\% quantile of $(0.30, 0.73)$.  Only three afterglows in the well-fit sample, and thirteen in the full sample, had an observer angle less than $0.2 \thO$.  Therefore, we find that most GRB afterglows are observed off-axis, at a significant fraction of the jet opening angle.  Off-axis viewing can have profound effects on the expected behaviour of the afterglow light curve, delaying and smoothing jet breaks out of the typical \swift{} observing window, which may contribute to the ``missing jet-break problem''.

Our sample include a number of short bursts. Although an orientation was obtained for GRB 051221A, the atypical nature of the light curve decay (including a  possible episode of energy injection, or multi-component jet), renders the application of our decelerating blast wave model questionable in this individual case. Constraining the orientation of short GRBs remains highly desirable given their potential as observable GW counterparts.  The other bursts for which opening angle and orientation are constrained do not share atypical features in the data preceding the regular decay stage.

The full results for this analysis, including the MCMC samples, will be made available on the Afterglow Library.


\section{Acknowledgements}

This research was supported in part by NASA through grant NNX10AF62G issued through the Astrophysics Theory Program, by the NSF through grant AST-1009863, by the Chandra grant TM3-14005X, and by Fermi grant NNX13AO93G.  HvE acknowledges support by the Alexander von Humboldt foundation.  BZ acknowledges the support of SAO contract SV4-74018, NASA contract NAS5-00136, and by SAO grants AR3-14005X, GO1-12102X, and GO3-14067X.  Resources supporting this work were provided by the NASA High-End Computing (HEC) Program through the NASA Advanced Supercomputing (NAS) Division at Ames Research Center. The software used in this work was in part developed by the DOE-supported ASCI/Alliance Center for Astrophysical Thermonuclear Flashes at the University of Chicago.  We thank Judith Racusin, David N. Burrows, David Hogg, and Daniel Foreman-Mackey for their many helpful discussions and comments.

\newpage


\bibliography{scalefit_sources}


\begin{appendix}
\section{Results For All Bursts}

\LongTables

\begin{deluxetable}{cccccc}
\tablecaption{Median Values of Non-Degenerate Parameters For All Bursts \label{tb:results}}
\tablewidth{0pt}
\tablehead{
\colhead{GRB} & \colhead{$\thO$} & \colhead{$\thobs/\thO$} & \colhead{$p$} & \colhead{Minimum $\chi^2/\text{dof}$} & \colhead{Extrapolation Used}}
\startdata
050126 & $0.377^{+0.086}_{-0.120}$ & $0.43^{+0.51}_{-0.31}$ & $2.26^{+0.10}_{-0.10}$ & $22.3/4$ & NoEx\\[2pt] 
050315\tablenotemark{a,b} & $0.343^{+0.038}_{-0.035}$ & $0.176^{+0.081}_{-0.099}$ & $2.084^{+0.030}_{-0.038}$ & $145.8/177$ & Ex\\[2pt] 
050318 & $0.145^{+0.097}_{-0.073}$ & $0.338^{+0.214}_{-0.089}$ & $2.30^{+0.22}_{-0.18}$ & $64.7/71$ & Ex\\[2pt] 
050319 & $0.0469^{+0.0052}_{-0.0014}$ & $0.627^{+0.054}_{-0.065}$ & $2.0103^{+0.0091}_{-0.0049}$ & $114.9/86$ & Ex\\[2pt] 
050401\tablenotemark{b} & $0.458^{+0.032}_{-0.069}$ & $0.71^{+0.26}_{-0.50}$ & $2.54^{+0.12}_{-0.12}$ & $359.6/316$ & NoEx\\[2pt] 
050408\tablenotemark{b} & $0.29^{+0.14}_{-0.16}$ & $0.54^{+0.29}_{-0.36}$ & $2.183^{+0.085}_{-0.084}$ & $35.0/43$ & Ex\\[2pt] 
050416A & $0.237^{+0.114}_{-0.059}$ & $0.34^{+0.24}_{-0.22}$ & $2.0058^{+0.0065}_{-0.0036}$ & $106.5/96$ & Ex\\[2pt] 
050505 & $0.19^{+0.11}_{-0.14}$ & $0.434^{+0.073}_{-0.114}$ & $2.32^{+0.12}_{-0.19}$ & $169.4/166$ & NoEx\\[2pt] 
050525A\tablenotemark{a,b} & $0.0551^{+0.0069}_{-0.0062}$ & $0.579^{+0.047}_{-0.052}$ & $2.044^{+0.041}_{-0.028}$ & $47.9/27$ & Ex\\[2pt] 
050603\tablenotemark{b} & $0.27^{+0.16}_{-0.15}$ & $0.53^{+0.30}_{-0.36}$ & $2.92^{+0.19}_{-0.24}$ & $49.9/48$ & NoEx\\[2pt] 
050730 & $0.420^{+0.056}_{-0.114}$ & $0.747^{+0.039}_{-0.142}$ & $3.719^{+0.084}_{-0.094}$ & $446.9/333$ & NoEx\\[2pt] 
050802 & $0.29^{+0.15}_{-0.15}$ & $0.55^{+0.32}_{-0.38}$ & $2.42^{+0.24}_{-0.20}$ & $23.2/24$ & Ex\\[2pt] 
050820A\tablenotemark{a} & $0.152^{+0.056}_{-0.022}$ & $0.57^{+0.13}_{-0.12}$ & $2.090^{+0.036}_{-0.032}$ & $358.5/321$ & NoEx\\[2pt] 
050824\tablenotemark{b} & $0.140^{+0.229}_{-0.088}$ & $0.39^{+0.30}_{-0.26}$ & $2.0143^{+0.0157}_{-0.0080}$ & $69.4/34$ & Ex\\[2pt] 
050826 & $0.354^{+0.099}_{-0.111}$ & $0.38^{+0.28}_{-0.26}$ & $2.092^{+0.176}_{-0.055}$ & $12.4/15$ & NoEx\\[2pt] 
050904 & $0.148^{+0.275}_{-0.095}$ & $0.48^{+0.20}_{-0.31}$ & $2.63^{+1.71}_{-0.55}$ & $28.2/11$ & Ex\\[2pt] 
050908 & $0.25^{+0.14}_{-0.11}$ & $0.33^{+0.28}_{-0.23}$ & $2.134^{+0.075}_{-0.075}$ & $33.1/7$ & NoEx\\[2pt] 
050922C\tablenotemark{a,b} & $0.074^{+0.033}_{-0.011}$ & $0.673^{+0.064}_{-0.075}$ & $2.100^{+0.228}_{-0.040}$ & $140.5/137$ & Ex\\[2pt] 
051016B\tablenotemark{b} & $0.34^{+0.12}_{-0.24}$ & $0.60^{+0.26}_{-0.20}$ & $2.085^{+0.085}_{-0.050}$ & $46.4/55$ & NoEx\\[2pt] 
051022 & $0.21^{+0.14}_{-0.13}$ & $0.14^{+0.36}_{-0.11}$ & $2.133^{+0.133}_{-0.080}$ & $52.8/51$ & NoEx\\[2pt] 
051109A & $0.29^{+0.14}_{-0.13}$ & $0.57^{+0.31}_{-0.39}$ & $2.21^{+0.22}_{-0.11}$ & $171.4/158$ & NoEx\\[2pt] 
051111\tablenotemark{b} & $0.28^{+0.15}_{-0.15}$ & $0.49^{+0.34}_{-0.33}$ & $2.78^{+0.35}_{-0.21}$ & $18.9/25$ & Ex\\[2pt] 
051221A\tablenotemark{a} & $0.448^{+0.031}_{-0.038}$ & $0.449^{+0.053}_{-0.058}$ & $2.024^{+0.025}_{-0.014}$ & $55.0/44$ & NoEx\\[2pt] 
060115 & $0.18^{+0.20}_{-0.11}$ & $0.76^{+0.16}_{-0.42}$ & $2.080^{+0.076}_{-0.053}$ & $31.5/21$ & Ex\\[2pt] 
060124\tablenotemark{b} & $0.21^{+0.20}_{-0.10}$ & $0.56^{+0.29}_{-0.32}$ & $2.40^{+0.17}_{-0.19}$ & $458.3/291$ & NoEx\\[2pt] 
060206\tablenotemark{b} & $0.377^{+0.084}_{-0.111}$ & $0.32^{+0.32}_{-0.23}$ & $2.089^{+0.103}_{-0.060}$ & $23.5/13$ & Ex\\[2pt] 
060210\tablenotemark{a} & $0.0604^{+0.0236}_{-0.0093}$ & $0.735^{+0.044}_{-0.117}$ & $2.111^{+0.047}_{-0.044}$ & $245.8/250$ & Ex\\[2pt] 
060218 & $0.33^{+0.12}_{-0.15}$ & $0.59^{+0.27}_{-0.38}$ & $2.24^{+0.17}_{-0.16}$ & $23.0/31$ & NoEx\\[2pt] 
060306 & $0.084^{+0.185}_{-0.029}$ & $0.78^{+0.13}_{-0.11}$ & $2.0141^{+0.0155}_{-0.0076}$ & $93.0/79$ & Ex\\[2pt] 
060418 & $0.31^{+0.13}_{-0.14}$ & $0.46^{+0.37}_{-0.32}$ & $2.70^{+0.13}_{-0.10}$ & $119.0/110$ & Ex\\[2pt] 
060502A & $0.32^{+0.12}_{-0.13}$ & $0.45^{+0.27}_{-0.30}$ & $2.0168^{+0.0149}_{-0.0094}$ & $50.1/60$ & Ex\\[2pt] 
060512\tablenotemark{b} & $0.27^{+0.15}_{-0.12}$ & $0.55^{+0.27}_{-0.35}$ & $2.116^{+0.108}_{-0.080}$ & $8.6/12$ & NoEx\\[2pt] 
060522\tablenotemark{a,b} & $0.326^{+0.113}_{-0.091}$ & $0.38^{+0.23}_{-0.24}$ & $2.099^{+0.062}_{-0.048}$ & $25.2/13$ & NoEx\\[2pt] 
060526 & $0.0468^{+0.0058}_{-0.0014}$ & $0.122^{+0.448}_{-0.087}$ & $2.023^{+0.047}_{-0.013}$ & $54.0/22$ & Ex\\[2pt] 
060604 & $0.059^{+0.071}_{-0.011}$ & $0.779^{+0.099}_{-0.082}$ & $2.043^{+0.041}_{-0.023}$ & $77.8/58$ & Ex\\[2pt] 
060605\tablenotemark{b} & $0.04614^{+0.18939}_{-0.00096}$ & $0.16^{+0.74}_{-0.11}$ & $2.0125^{+1.4031}_{-0.0087}$ & $128.2/69$ & Ex\\[2pt] 
060607A\tablenotemark{a,b} & $0.374^{+0.095}_{-0.072}$ & $0.593^{+0.078}_{-0.098}$ & $4.63^{+0.17}_{-0.17}$ & $257.2/156$ & Ex\\[2pt] 
060614\tablenotemark{b} & $0.293^{+0.122}_{-0.085}$ & $0.38^{+0.43}_{-0.22}$ & $2.100^{+0.101}_{-0.064}$ & $54.8/79$ & Ex\\[2pt] 
060707\tablenotemark{b} & $0.22^{+0.19}_{-0.16}$ & $0.59^{+0.28}_{-0.38}$ & $2.135^{+0.048}_{-0.069}$ & $41.8/21$ & Ex\\[2pt] 
060714\tablenotemark{a} & $0.0557^{+0.0104}_{-0.0072}$ & $0.743^{+0.057}_{-0.050}$ & $2.058^{+0.060}_{-0.035}$ & $73.3/43$ & Ex\\[2pt] 
060729 & $0.370^{+0.047}_{-0.039}$ & $0.636^{+0.030}_{-0.036}$ & $2.079^{+0.039}_{-0.033}$ & $584.3/561$ & Ex\\[2pt] 
060801\tablenotemark{a} & $0.0562^{+0.0056}_{-0.0064}$ & $0.050^{+0.055}_{-0.035}$ & $4.84^{+0.12}_{-0.22}$ & $21.1/8$ & NoEx\\[2pt] 
060814\tablenotemark{b} & $0.360^{+0.098}_{-0.123}$ & $0.36^{+0.38}_{-0.25}$ & $2.108^{+0.145}_{-0.075}$ & $10.1/12$ & NoEx\\[2pt] 
060904B\tablenotemark{a,b} & $0.083^{+0.048}_{-0.015}$ & $0.734^{+0.103}_{-0.078}$ & $2.114^{+0.047}_{-0.056}$ & $50.8/49$ & Ex\\[2pt] 
060906 & $0.0500^{+0.0120}_{-0.0042}$ & $0.620^{+0.080}_{-0.076}$ & $2.0115^{+0.0135}_{-0.0067}$ & $166.6/31$ & Ex\\[2pt] 
060908\tablenotemark{b} & $0.384^{+0.082}_{-0.114}$ & $0.38^{+0.42}_{-0.27}$ & $2.76^{+0.14}_{-0.14}$ & $54.1/27$ & NoEx\\[2pt] 
060912A\tablenotemark{b} & $0.213^{+0.123}_{-0.043}$ & $0.59^{+0.19}_{-0.36}$ & $2.148^{+0.070}_{-0.083}$ & $33.2/30$ & NoEx\\[2pt] 
060926 & $0.28^{+0.15}_{-0.15}$ & $0.51^{+0.33}_{-0.35}$ & $2.43^{+0.26}_{-0.22}$ & $3.7/3$ & NoEx\\[2pt] 
060927\tablenotemark{b} & $0.31^{+0.13}_{-0.15}$ & $0.51^{+0.34}_{-0.35}$ & $3.81^{+0.76}_{-0.74}$ & $9.8/10$ & NoEx\\[2pt] 
061007\tablenotemark{b} & $0.32^{+0.13}_{-0.17}$ & $0.52^{+0.29}_{-0.30}$ & $2.48^{+0.26}_{-0.26}$ & $9.4/9$ & NoEx\\[2pt] 
061021\tablenotemark{b} & $0.133^{+0.060}_{-0.021}$ & $0.56^{+0.17}_{-0.14}$ & $2.035^{+0.032}_{-0.023}$ & $284.8/290$ & Ex\\[2pt] 
061121\tablenotemark{b} & $0.094^{+0.118}_{-0.030}$ & $0.858^{+0.079}_{-0.075}$ & $2.474^{+0.087}_{-0.088}$ & $144.3/163$ & Ex\\[2pt] 
061222A\tablenotemark{a,b} & $0.0695^{+0.0102}_{-0.0053}$ & $0.580^{+0.048}_{-0.034}$ & $2.441^{+0.049}_{-0.047}$ & $387.3/277$ & NoEx\\[2pt] 
070110\tablenotemark{b} & $0.33^{+0.12}_{-0.15}$ & $0.43^{+0.33}_{-0.29}$ & $2.130^{+0.154}_{-0.090}$ & $25.1/18$ & Ex\\[2pt] 
070125 & $0.26^{+0.12}_{-0.11}$ & $0.41^{+0.23}_{-0.14}$ & $2.36^{+0.28}_{-0.22}$ & $38.9/38$ & NoEx\\[2pt] 
070208 & $0.28^{+0.15}_{-0.15}$ & $0.54^{+0.31}_{-0.36}$ & $2.32^{+0.35}_{-0.19}$ & $20.1/13$ & Ex\\[2pt] 
070306\tablenotemark{a,b} & $0.291^{+0.030}_{-0.037}$ & $0.361^{+0.044}_{-0.046}$ & $2.082^{+0.083}_{-0.043}$ & $137.5/102$ & NoEx\\[2pt] 
070318\tablenotemark{b} & $0.30^{+0.11}_{-0.11}$ & $0.23^{+0.39}_{-0.16}$ & $2.094^{+0.173}_{-0.066}$ & $96.6/54$ & NoEx\\[2pt] 
070411 & $0.33^{+0.12}_{-0.16}$ & $0.58^{+0.21}_{-0.38}$ & $2.116^{+0.191}_{-0.075}$ & $21.2/23$ & Ex\\[2pt] 
070419A\tablenotemark{b} & $0.18^{+0.19}_{-0.11}$ & $0.65^{+0.20}_{-0.32}$ & $4.884^{+0.089}_{-0.208}$ & $138.9/103$ & Ex\\[2pt] 
070506\tablenotemark{b} & $0.25^{+0.17}_{-0.16}$ & $0.50^{+0.31}_{-0.34}$ & $2.044^{+0.054}_{-0.028}$ & $10.9/4$ & Ex\\[2pt] 
070508\tablenotemark{b} & $0.449^{+0.040}_{-0.081}$ & $0.768^{+0.053}_{-0.616}$ & $2.549^{+0.190}_{-0.052}$ & $466.7/487$ & Ex\\[2pt] 
070521\tablenotemark{b} & $0.145^{+0.152}_{-0.075}$ & $0.42^{+0.31}_{-0.18}$ & $2.29^{+0.33}_{-0.19}$ & $64.6/65$ & NoEx\\[2pt] 
070529 & $0.172^{+0.152}_{-0.054}$ & $0.786^{+0.087}_{-0.129}$ & $2.185^{+0.066}_{-0.074}$ & $26.1/26$ & NoEx\\[2pt] 
070611\tablenotemark{b} & $0.27^{+0.15}_{-0.11}$ & $0.41^{+0.36}_{-0.29}$ & $2.19^{+0.22}_{-0.13}$ & $15.9/4$ & Ex\\[2pt] 
070714B\tablenotemark{b} & $0.33^{+0.11}_{-0.11}$ & $0.835^{+0.059}_{-0.617}$ & $2.670^{+0.212}_{-0.073}$ & $114.7/67$ & Ex\\[2pt] 
070721B & $0.081^{+0.036}_{-0.027}$ & $0.125^{+0.099}_{-0.085}$ & $2.129^{+0.104}_{-0.070}$ & $45.9/51$ & NoEx\\[2pt] 
070724A\tablenotemark{b} & $0.28^{+0.15}_{-0.16}$ & $0.51^{+0.34}_{-0.35}$ & $4.52^{+0.29}_{-0.36}$ & $82.9/3$ & NoEx\\[2pt] 
070802\tablenotemark{b} & $0.477^{+0.018}_{-0.039}$ & $0.31^{+0.19}_{-0.19}$ & $2.055^{+0.027}_{-0.022}$ & $72.0/8$ & NoEx\\[2pt] 
070810A\tablenotemark{b} & $0.104^{+0.196}_{-0.049}$ & $0.76^{+0.16}_{-0.29}$ & $2.105^{+0.093}_{-0.066}$ & $25.2/27$ & Ex\\[2pt] 
071003\tablenotemark{b} & $0.30^{+0.14}_{-0.18}$ & $0.61^{+0.25}_{-0.33}$ & $2.85^{+0.56}_{-0.37}$ & $66.0/59$ & Ex\\[2pt] 
071010A\tablenotemark{b} & $0.30^{+0.13}_{-0.15}$ & $0.54^{+0.31}_{-0.35}$ & $2.97^{+0.32}_{-0.36}$ & $2.6/3$ & Ex\\[2pt] 
071010B\tablenotemark{b} & $0.34^{+0.12}_{-0.22}$ & $0.33^{+0.37}_{-0.23}$ & $2.036^{+0.039}_{-0.023}$ & $7.0/8$ & Ex\\[2pt] 
071020\tablenotemark{b} & $0.213^{+0.238}_{-0.026}$ & $0.872^{+0.021}_{-0.405}$ & $2.114^{+0.049}_{-0.020}$ & $252.4/185$ & NoEx\\[2pt] 
071031\tablenotemark{b} & $0.32^{+0.12}_{-0.14}$ & $0.41^{+0.35}_{-0.27}$ & $2.119^{+0.132}_{-0.085}$ & $10.2/3$ & NoEx\\[2pt] 
071112C\tablenotemark{b} & $0.394^{+0.073}_{-0.117}$ & $0.79^{+0.15}_{-0.66}$ & $2.78^{+0.20}_{-0.17}$ & $27.3/23$ & NoEx\\[2pt] 
071117\tablenotemark{b} & $0.30^{+0.14}_{-0.16}$ & $0.45^{+0.32}_{-0.29}$ & $2.068^{+0.080}_{-0.045}$ & $29.8/21$ & NoEx\\[2pt] 
071122\tablenotemark{b} & $0.25^{+0.16}_{-0.16}$ & $0.53^{+0.34}_{-0.35}$ & $4.30^{+0.52}_{-0.42}$ & $39.5/29$ & Ex\\[2pt] 
080210\tablenotemark{a,b} & $0.223^{+0.080}_{-0.142}$ & $0.38^{+0.13}_{-0.27}$ & $2.093^{+0.130}_{-0.057}$ & $56.9/25$ & NoEx\\[2pt] 
080310\tablenotemark{b} & $0.097^{+0.263}_{-0.052}$ & $0.30^{+0.46}_{-0.16}$ & $2.89^{+0.59}_{-0.88}$ & $57.9/33$ & Ex\\[2pt] 
080319B\tablenotemark{a,b} & $0.098^{+0.046}_{-0.013}$ & $0.620^{+0.082}_{-0.080}$ & $2.780^{+0.018}_{-0.012}$ & $1630.1/1585$ & Ex\\[2pt] 
080319C & $0.30^{+0.14}_{-0.15}$ & $0.46^{+0.35}_{-0.32}$ & $2.80^{+0.11}_{-0.11}$ & $44.2/46$ & Ex\\[2pt] 
080411\tablenotemark{b} & $0.097^{+0.362}_{-0.017}$ & $0.771^{+0.034}_{-0.043}$ & $2.097^{+0.093}_{-0.046}$ & $375.7/395$ & Ex\\[2pt] 
080413A & $0.26^{+0.17}_{-0.16}$ & $0.52^{+0.32}_{-0.33}$ & $2.27^{+0.23}_{-0.14}$ & $7.1/5$ & NoEx\\[2pt] 
080413B\tablenotemark{b} & $0.1350^{+0.0060}_{-0.0055}$ & $0.369^{+0.043}_{-0.046}$ & $2.0106^{+0.0090}_{-0.0052}$ & $316.9/229$ & NoEx\\[2pt] 
080430\tablenotemark{a} & $0.0553^{+0.0052}_{-0.0079}$ & $0.587^{+0.060}_{-0.067}$ & $2.0026^{+0.0029}_{-0.0013}$ & $274.1/138$ & Ex\\[2pt] 
080605\tablenotemark{a,b} & $0.355^{+0.090}_{-0.100}$ & $0.50^{+0.18}_{-0.25}$ & $2.537^{+0.071}_{-0.167}$ & $327.9/308$ & NoEx\\[2pt] 
080607\tablenotemark{b} & $0.26^{+0.16}_{-0.14}$ & $0.54^{+0.30}_{-0.33}$ & $2.65^{+0.39}_{-0.23}$ & $27.9/24$ & Ex\\[2pt] 
080707 & $0.30^{+0.14}_{-0.15}$ & $0.44^{+0.35}_{-0.30}$ & $2.21^{+0.26}_{-0.14}$ & $26.7/10$ & Ex\\[2pt] 
080710 & $0.244^{+0.092}_{-0.130}$ & $0.20^{+0.15}_{-0.12}$ & $2.089^{+0.107}_{-0.050}$ & $82.8/60$ & Ex\\[2pt] 
080721\tablenotemark{a,b} & $0.1117^{+0.0109}_{-0.0083}$ & $0.731^{+0.025}_{-0.021}$ & $2.397^{+0.016}_{-0.014}$ & $1485.3/1382$ & Ex\\[2pt] 
080804 & $0.297^{+0.122}_{-0.093}$ & $0.68^{+0.12}_{-0.17}$ & $2.039^{+0.026}_{-0.019}$ & $86.4/95$ & Ex\\[2pt] 
080805 & $0.31^{+0.13}_{-0.14}$ & $0.39^{+0.30}_{-0.27}$ & $2.094^{+0.189}_{-0.071}$ & $18.5/12$ & Ex\\[2pt] 
080810\tablenotemark{b} & $0.34^{+0.11}_{-0.27}$ & $0.41^{+0.28}_{-0.33}$ & $2.63^{+0.17}_{-0.29}$ & $66.7/70$ & Ex\\[2pt] 
080905A & $0.27^{+0.15}_{-0.15}$ & $0.49^{+0.36}_{-0.34}$ & $3.80^{+0.40}_{-0.34}$ & $17.4/5$ & NoEx\\[2pt] 
080905B & $0.161^{+0.179}_{-0.050}$ & $0.58^{+0.22}_{-0.43}$ & $2.46^{+0.13}_{-0.13}$ & $60.8/57$ & Ex\\[2pt] 
080913\tablenotemark{b} & $0.359^{+0.099}_{-0.125}$ & $0.40^{+0.42}_{-0.28}$ & $2.21^{+0.15}_{-0.11}$ & $42.7/4$ & Ex\\[2pt] 
080916A & $0.156^{+0.222}_{-0.086}$ & $0.78^{+0.13}_{-0.28}$ & $2.099^{+0.090}_{-0.062}$ & $47.3/42$ & Ex\\[2pt] 
080928\tablenotemark{b} & $0.25^{+0.16}_{-0.17}$ & $0.62^{+0.26}_{-0.33}$ & $3.17^{+0.36}_{-0.56}$ & $83.2/68$ & Ex\\[2pt] 
081007\tablenotemark{b} & $0.159^{+0.183}_{-0.066}$ & $0.76^{+0.13}_{-0.19}$ & $2.046^{+0.029}_{-0.025}$ & $65.8/55$ & Ex\\[2pt] 
081008 & $0.0610^{+0.0092}_{-0.0089}$ & $0.34^{+0.10}_{-0.14}$ & $2.088^{+0.066}_{-0.056}$ & $40.7/46$ & Ex\\[2pt] 
081028\tablenotemark{b} & $0.30^{+0.13}_{-0.15}$ & $0.53^{+0.32}_{-0.36}$ & $2.92^{+0.33}_{-0.34}$ & $22.7/4$ & Ex\\[2pt] 
081029\tablenotemark{b} & $0.1615^{+0.0055}_{-0.0061}$ & $0.031^{+0.034}_{-0.022}$ & $2.075^{+0.025}_{-0.021}$ & $140.2/77$ & Ex\\[2pt] 
081121 & $0.281^{+0.143}_{-0.092}$ & $0.760^{+0.096}_{-0.351}$ & $2.463^{+0.102}_{-0.096}$ & $141.0/139$ & Ex\\[2pt] 
081203A & $0.150^{+0.034}_{-0.072}$ & $0.411^{+0.056}_{-0.065}$ & $2.113^{+0.083}_{-0.056}$ & $214.1/216$ & NoEx\\[2pt] 
081221\tablenotemark{b} & $0.343^{+0.108}_{-0.094}$ & $0.33^{+0.22}_{-0.22}$ & $2.406^{+0.031}_{-0.032}$ & $248.0/255$ & Ex\\[2pt] 
081222\tablenotemark{b} & $0.0844^{+0.0081}_{-0.0120}$ & $0.275^{+0.099}_{-0.114}$ & $2.408^{+0.032}_{-0.297}$ & $410.6/398$ & Ex\\[2pt] 
090102\tablenotemark{b} & $0.382^{+0.082}_{-0.098}$ & $0.75^{+0.13}_{-0.53}$ & $2.544^{+0.154}_{-0.067}$ & $161.6/134$ & NoEx\\[2pt] 
090205\tablenotemark{a} & $0.0511^{+0.0079}_{-0.0045}$ & $0.21^{+0.12}_{-0.13}$ & $2.042^{+0.047}_{-0.026}$ & $23.4/20$ & NoEx\\[2pt] 
090313\tablenotemark{b} & $0.127^{+0.143}_{-0.040}$ & $0.15^{+0.39}_{-0.11}$ & $2.151^{+0.479}_{-0.097}$ & $37.2/40$ & Ex\\[2pt] 
090323\tablenotemark{b} & $0.31^{+0.13}_{-0.15}$ & $0.53^{+0.32}_{-0.37}$ & $2.77^{+0.22}_{-0.22}$ & $16.5/10$ & NoEx\\[2pt] 
090328A\tablenotemark{b} & $0.31^{+0.13}_{-0.16}$ & $0.56^{+0.29}_{-0.36}$ & $2.76^{+0.28}_{-0.29}$ & $12.4/9$ & NoEx\\[2pt] 
090418A\tablenotemark{b} & $0.0686^{+0.4036}_{-0.0048}$ & $0.314^{+0.067}_{-0.115}$ & $2.0037^{+0.4857}_{-0.0020}$ & $100.8/108$ & NoEx\\[2pt] 
090423 & $0.26^{+0.16}_{-0.14}$ & $0.51^{+0.32}_{-0.35}$ & $2.55^{+0.31}_{-0.16}$ & $16.1/23$ & Ex\\[2pt] 
090424\tablenotemark{a,b} & $0.215^{+0.017}_{-0.014}$ & $0.761^{+0.018}_{-0.020}$ & $2.026^{+0.019}_{-0.012}$ & $568.3/537$ & NoEx\\[2pt] 
090426\tablenotemark{b} & $0.31^{+0.13}_{-0.14}$ & $0.48^{+0.32}_{-0.33}$ & $2.220^{+0.091}_{-0.117}$ & $27.6/22$ & Ex\\[2pt] 
090510\tablenotemark{b} & $0.387^{+0.080}_{-0.106}$ & $0.57^{+0.22}_{-0.39}$ & $3.34^{+0.38}_{-0.29}$ & $74.8/66$ & Ex\\[2pt] 
090516\tablenotemark{a} & $0.0655^{+0.0035}_{-0.0047}$ & $0.320^{+0.039}_{-0.039}$ & $2.046^{+0.038}_{-0.024}$ & $144.7/133$ & NoEx\\[2pt] 
090519\tablenotemark{b} & $0.27^{+0.15}_{-0.16}$ & $0.54^{+0.37}_{-0.37}$ & $4.40^{+0.28}_{-0.30}$ & $119.1/15$ & NoEx\\[2pt] 
090529 & $0.477^{+0.017}_{-0.029}$ & $0.141^{+0.139}_{-0.098}$ & $2.019^{+0.017}_{-0.011}$ & $16.6/5$ & NoEx\\[2pt] 
090618\tablenotemark{a,b} & $0.0590^{+0.0025}_{-0.0026}$ & $0.758^{+0.011}_{-0.012}$ & $2.224^{+0.018}_{-0.017}$ & $979.1/843$ & Ex\\[2pt] 
090709A\tablenotemark{a} & $0.279^{+0.126}_{-0.084}$ & $0.72^{+0.10}_{-0.28}$ & $2.492^{+0.087}_{-0.074}$ & $167.6/170$ & NoEx\\[2pt] 
090715B\tablenotemark{b} & $0.31^{+0.12}_{-0.12}$ & $0.50^{+0.24}_{-0.23}$ & $2.132^{+0.158}_{-0.084}$ & $38.0/25$ & NoEx\\[2pt] 
090726\tablenotemark{b} & $0.170^{+0.195}_{-0.084}$ & $0.44^{+0.39}_{-0.32}$ & $2.42^{+0.22}_{-0.23}$ & $16.5/22$ & NoEx\\[2pt] 
090809 & $0.31^{+0.12}_{-0.11}$ & $0.49^{+0.18}_{-0.21}$ & $2.052^{+0.106}_{-0.034}$ & $22.6/16$ & NoEx\\[2pt] 
090812\tablenotemark{b} & $0.29^{+0.15}_{-0.13}$ & $0.69^{+0.20}_{-0.44}$ & $2.23^{+0.11}_{-0.12}$ & $39.6/41$ & NoEx\\[2pt] 
090902B\tablenotemark{b} & $0.32^{+0.13}_{-0.16}$ & $0.58^{+0.27}_{-0.39}$ & $2.45^{+0.19}_{-0.18}$ & $65.5/70$ & NoEx\\[2pt] 
090926A\tablenotemark{b} & $0.32^{+0.13}_{-0.16}$ & $0.61^{+0.26}_{-0.40}$ & $2.50^{+0.22}_{-0.20}$ & $90.9/64$ & NoEx\\[2pt] 
090926B & $0.35^{+0.11}_{-0.12}$ & $0.39^{+0.37}_{-0.27}$ & $2.164^{+0.135}_{-0.093}$ & $22.3/18$ & Ex\\[2pt] 
090927\tablenotemark{b} & $0.448^{+0.038}_{-0.131}$ & $0.59^{+0.11}_{-0.16}$ & $2.048^{+0.048}_{-0.029}$ & $32.0/21$ & NoEx\\[2pt] 
091003\tablenotemark{b} & $0.31^{+0.13}_{-0.17}$ & $0.58^{+0.28}_{-0.39}$ & $2.39^{+0.21}_{-0.21}$ & $43.8/39$ & Ex\\[2pt] 
091018\tablenotemark{b} & $0.30^{+0.14}_{-0.15}$ & $0.57^{+0.29}_{-0.38}$ & $2.69^{+0.22}_{-0.20}$ & $40.1/33$ & NoEx\\[2pt] 
091020\tablenotemark{b} & $0.28^{+0.15}_{-0.13}$ & $0.65^{+0.19}_{-0.39}$ & $2.426^{+0.066}_{-0.078}$ & $180.9/185$ & Ex\\[2pt] 
091024\tablenotemark{b} & $0.31^{+0.13}_{-0.15}$ & $0.50^{+0.36}_{-0.34}$ & $3.24^{+0.28}_{-0.18}$ & $107.2/19$ & Ex\\[2pt] 
091029\tablenotemark{b} & $0.133^{+0.252}_{-0.038}$ & $0.73^{+0.15}_{-0.21}$ & $2.065^{+0.058}_{-0.037}$ & $110.5/105$ & NoEx\\[2pt] 
091109A & $0.221^{+0.161}_{-0.068}$ & $0.44^{+0.26}_{-0.28}$ & $2.061^{+0.053}_{-0.040}$ & $14.0/17$ & NoEx\\[2pt] 
091127\tablenotemark{b} & $0.151^{+0.193}_{-0.071}$ & $0.905^{+0.049}_{-0.078}$ & $2.549^{+0.078}_{-0.098}$ & $354.7/363$ & Ex\\[2pt] 
091208B\tablenotemark{a} & $0.0952^{+0.0098}_{-0.0127}$ & $0.19^{+0.25}_{-0.13}$ & $2.082^{+0.056}_{-0.047}$ & $75.4/57$ & Ex\\[2pt] 
100117A\tablenotemark{b} & $0.28^{+0.15}_{-0.16}$ & $0.51^{+0.34}_{-0.35}$ & $3.07^{+0.57}_{-0.50}$ & $32.1/9$ & NoEx\\[2pt] 
100219A & $0.0481^{+0.0052}_{-0.0024}$ & $0.064^{+0.072}_{-0.045}$ & $2.0118^{+0.0138}_{-0.0071}$ & $66.7/20$ & Ex\\[2pt] 
100302A\tablenotemark{b} & $0.19^{+0.20}_{-0.13}$ & $0.37^{+0.31}_{-0.25}$ & $2.0156^{+0.0189}_{-0.0096}$ & $26.3/18$ & Ex\\[2pt] 
100316B & $0.151^{+0.226}_{-0.090}$ & $0.79^{+0.14}_{-0.34}$ & $2.079^{+0.082}_{-0.050}$ & $38.7/7$ & Ex\\[2pt] 
100418A\tablenotemark{b} & $0.24^{+0.16}_{-0.11}$ & $0.20^{+0.41}_{-0.15}$ & $2.18^{+0.25}_{-0.12}$ & $18.8/11$ & NoEx\\[2pt] 
100424A & $0.0552^{+0.0852}_{-0.0081}$ & $0.54^{+0.21}_{-0.28}$ & $3.32^{+0.27}_{-0.21}$ & $189.3/156$ & NoEx\\[2pt] 
100425A\tablenotemark{b} & $0.16^{+0.22}_{-0.10}$ & $0.45^{+0.29}_{-0.31}$ & $2.016^{+0.022}_{-0.010}$ & $38.0/16$ & Ex\\[2pt] 
100513A\tablenotemark{b} & $0.28^{+0.15}_{-0.15}$ & $0.45^{+0.32}_{-0.31}$ & $2.170^{+0.067}_{-0.085}$ & $39.5/20$ & Ex\\[2pt] 
100615A & $0.28^{+0.15}_{-0.17}$ & $0.48^{+0.34}_{-0.34}$ & $2.16^{+0.21}_{-0.11}$ & $51.3/55$ & Ex\\[2pt] 
100621A\tablenotemark{b} & $0.0477^{+0.0047}_{-0.0020}$ & $0.583^{+0.049}_{-0.047}$ & $2.075^{+0.030}_{-0.029}$ & $212.1/181$ & Ex\\[2pt] 
100724A\tablenotemark{b} & $0.28^{+0.15}_{-0.15}$ & $0.51^{+0.31}_{-0.34}$ & $2.25^{+0.12}_{-0.19}$ & $16.7/11$ & Ex\\[2pt] 
100728A\tablenotemark{a,b} & $0.112^{+0.052}_{-0.011}$ & $0.722^{+0.119}_{-0.062}$ & $2.438^{+0.073}_{-0.055}$ & $253.2/249$ & Ex\\[2pt] 
100728B\tablenotemark{b} & $0.30^{+0.12}_{-0.19}$ & $0.56^{+0.11}_{-0.18}$ & $2.136^{+0.087}_{-0.058}$ & $47.6/28$ & NoEx\\[2pt] 
100816A\tablenotemark{b} & $0.29^{+0.14}_{-0.15}$ & $0.45^{+0.33}_{-0.31}$ & $2.22^{+0.21}_{-0.15}$ & $24.8/17$ & Ex\\[2pt] 
100901A\tablenotemark{a} & $0.413^{+0.033}_{-0.033}$ & $0.473^{+0.032}_{-0.031}$ & $2.045^{+0.039}_{-0.025}$ & $160.4/99$ & Ex\\[2pt] 
100906A\tablenotemark{a} & $0.0540^{+0.0038}_{-0.0079}$ & $0.307^{+0.031}_{-0.030}$ & $2.0126^{+0.0130}_{-0.0071}$ & $252.6/128$ & Ex\\[2pt] 
101219A\tablenotemark{b} & $0.29^{+0.14}_{-0.14}$ & $0.49^{+0.34}_{-0.34}$ & $3.82^{+0.28}_{-0.24}$ & $46.5/5$ & NoEx\\[2pt] 
110128A\tablenotemark{b} & $0.469^{+0.023}_{-0.047}$ & $0.15^{+0.19}_{-0.11}$ & $2.032^{+0.030}_{-0.020}$ & $34.3/8$ & Ex\\[2pt] 
110205A\tablenotemark{b} & $0.388^{+0.077}_{-0.145}$ & $0.69^{+0.25}_{-0.54}$ & $2.759^{+0.088}_{-0.134}$ & $142.1/164$ & Ex\\[2pt] 
110213A\tablenotemark{b} & $0.29^{+0.14}_{-0.15}$ & $0.53^{+0.31}_{-0.37}$ & $3.09^{+0.22}_{-0.19}$ & $74.4/79$ & Ex\\[2pt] 
110422A\tablenotemark{a,b} & $0.0733^{+0.0111}_{-0.0098}$ & $0.676^{+0.035}_{-0.050}$ & $2.285^{+0.049}_{-0.063}$ & $238.9/254$ & NoEx\\[2pt] 
110503A\tablenotemark{b} & $0.402^{+0.055}_{-0.284}$ & $0.734^{+0.056}_{-0.195}$ & $2.073^{+0.051}_{-0.056}$ & $402.2/389$ & NoEx\\[2pt] 
110715A & $0.34^{+0.11}_{-0.18}$ & $0.39^{+0.34}_{-0.27}$ & $2.085^{+0.075}_{-0.052}$ & $8.4/14$ & NoEx\\[2pt] 
110731A & $0.184^{+0.052}_{-0.035}$ & $0.716^{+0.110}_{-0.065}$ & $2.112^{+0.020}_{-0.019}$ & $275.4/274$ & Ex\\[2pt] 
110801A\tablenotemark{a,b} & $0.419^{+0.056}_{-0.075}$ & $0.50^{+0.13}_{-0.16}$ & $2.335^{+0.031}_{-0.041}$ & $168.8/118$ & Ex\\[2pt] 
110808A\tablenotemark{b} & $0.19^{+0.19}_{-0.13}$ & $0.33^{+0.31}_{-0.23}$ & $2.0129^{+0.0170}_{-0.0084}$ & $18.4/6$ & Ex\\[2pt] 
110818A & $0.23^{+0.16}_{-0.15}$ & $0.63^{+0.24}_{-0.31}$ & $2.25^{+0.14}_{-0.14}$ & $28.0/21$ & NoEx\\[2pt] 
111008A & $0.04612^{+0.00259}_{-0.00085}$ & $0.721^{+0.027}_{-0.027}$ & $2.028^{+0.019}_{-0.014}$ & $244.7/130$ & Ex\\[2pt] 
111107A\tablenotemark{b} & $0.28^{+0.15}_{-0.15}$ & $0.43^{+0.35}_{-0.29}$ & $2.19^{+0.13}_{-0.13}$ & $13.6/9$ & Ex\\[2pt] 
111209A & $0.34^{+0.11}_{-0.13}$ & $0.61^{+0.28}_{-0.40}$ & $2.61^{+0.13}_{-0.16}$ & $120.2/75$ & NoEx\\[2pt] 
111228A & $0.04591^{+0.00177}_{-0.00070}$ & $0.754^{+0.017}_{-0.021}$ & $2.0057^{+0.0058}_{-0.0033}$ & $304.5/155$ & Ex\\[2pt] 
111229A & $0.25^{+0.17}_{-0.15}$ & $0.52^{+0.33}_{-0.36}$ & $4.13^{+0.51}_{-0.56}$ & $10.6/5$ & Ex\\[2pt] 
120118B & $0.29^{+0.14}_{-0.15}$ & $0.44^{+0.34}_{-0.30}$ & $2.16^{+0.18}_{-0.10}$ & $16.1/17$ & NoEx\\[2pt] 
120119A\tablenotemark{b} & $0.0510^{+0.0481}_{-0.0046}$ & $0.36^{+0.17}_{-0.24}$ & $2.479^{+0.039}_{-0.377}$ & $140.9/87$ & Ex\\[2pt] 
120327A\tablenotemark{b} & $0.412^{+0.060}_{-0.088}$ & $0.29^{+0.31}_{-0.20}$ & $2.67^{+0.13}_{-0.18}$ & $87.8/60$ & NoEx\\[2pt] 
120404A & $0.28^{+0.15}_{-0.15}$ & $0.49^{+0.35}_{-0.34}$ & $3.02^{+0.30}_{-0.27}$ & $18.9/20$ & NoEx\\[2pt] 
120422A & $0.20^{+0.20}_{-0.14}$ & $0.43^{+0.31}_{-0.28}$ & $2.017^{+0.023}_{-0.012}$ & $45.8/4$ & Ex\\[2pt] 
120711A & $0.25^{+0.16}_{-0.11}$ & $0.65^{+0.19}_{-0.40}$ & $2.687^{+0.067}_{-0.140}$ & $258.2/239$ & Ex\\[2pt] 
120712A\tablenotemark{b} & $0.301^{+0.125}_{-0.099}$ & $0.878^{+0.039}_{-0.069}$ & $2.211^{+0.032}_{-0.030}$ & $67.6/56$ & Ex\\[2pt] 
120802A\tablenotemark{b} & $0.21^{+0.18}_{-0.15}$ & $0.55^{+0.28}_{-0.38}$ & $2.029^{+0.033}_{-0.017}$ & $104.1/18$ & Ex\\[2pt] 
120811C & $0.23^{+0.18}_{-0.14}$ & $0.65^{+0.27}_{-0.44}$ & $2.133^{+0.098}_{-0.077}$ & $41.2/42$ & Ex\\[2pt] 
120815A & $0.32^{+0.13}_{-0.16}$ & $0.41^{+0.34}_{-0.29}$ & $2.135^{+0.102}_{-0.088}$ & $41.5/38$ & NoEx\\[2pt] 
120907A\tablenotemark{a} & $0.1170^{+0.0098}_{-0.0073}$ & $0.292^{+0.096}_{-0.104}$ & $2.0139^{+0.0127}_{-0.0075}$ & $112.7/84$ & NoEx\\[2pt] 
120909A\tablenotemark{b} & $0.22^{+0.18}_{-0.13}$ & $0.69^{+0.21}_{-0.35}$ & $2.44^{+0.43}_{-0.23}$ & $121.3/115$ & Ex\\[2pt] 
121024A\tablenotemark{a} & $0.0565^{+0.0179}_{-0.0080}$ & $0.35^{+0.24}_{-0.22}$ & $2.142^{+0.093}_{-0.081}$ & $43.9/42$ & Ex\\[2pt] 
121027A & $0.058^{+0.076}_{-0.011}$ & $0.61^{+0.21}_{-0.10}$ & $2.116^{+0.054}_{-0.049}$ & $103.2/66$ & Ex\\[2pt] 
121128A & $0.27^{+0.15}_{-0.14}$ & $0.51^{+0.31}_{-0.35}$ & $2.98^{+0.15}_{-0.21}$ & $92.8/81$ & Ex\\[2pt] 
121201A & $0.364^{+0.098}_{-0.106}$ & $0.39^{+0.48}_{-0.27}$ & $2.269^{+0.053}_{-0.098}$ & $35.2/26$ & Ex\\[2pt] 
121211A & $0.30^{+0.15}_{-0.22}$ & $0.61^{+0.27}_{-0.32}$ & $2.093^{+0.143}_{-0.063}$ & $68.1/61$ & Ex\\[2pt] 
121229A & $0.28^{+0.15}_{-0.15}$ & $0.53^{+0.35}_{-0.36}$ & $4.12^{+0.55}_{-0.50}$ & $26.8/16$ & NoEx\\[2pt] 
\enddata
\tablenotetext{a}{Well-Fit in $\thO$ and $\thobs/\thO$.}
\tablenotetext{b}{Well-Fit in $p$.}

\tablecomments{The median values of the posterior distributions for each of $\thO$, $\thobs/\thO$, and $p$.  Uncertainties are given at the $68\%$ level. The last column denotes whether the fit reported extrapolated in time outside the tabulated values of $\cfp$, $\cfm$, and $\cfc$.}

\end{deluxetable} 

\end{appendix}

\end{document}